\documentclass[reqno,final]{amsart}

\usepackage{amssymb}

\usepackage{graphicx}

\usepackage{booktabs}
\usepackage{paralist}
\allowdisplaybreaks[1]

\numberwithin{equation}{section}

\begin{document}

\title[Oscillations in coupled CR pairs]{Awakened oscillations in coupled consumer-resource pairs}

\author[A. Mustafin]{Almaz Mustafin}
\address{Department of General and Theoretical Physics\\
Kazakh National Technical University\\
22 Satpayev St., Almaty, 050013, Kazakhstan}

\email{mustafin\_a1@kazntu.kz}

\keywords{antiphase oscillations, consumer-resource, coupled oscillators, heteroclinic cycles,  laser dynamics, predator-prey, rate equations, relaxation oscillations, synchronization}

\subjclass[2010]{34C15, 34C26, 78A60, 92B25, 92D25}

\date{}

\begin{abstract}
The paper concerns two interacting consumer-resource pairs based on che\-mo\-stat-like equations under the assumption that the dynamics of the resource is considerably slower than that of the consumer. The presence of two different time scales enables to carry out a fairly complete analysis of the problem. This is done by treating consumers and resources in the coupled system as fast-scale and slow-scale variables respectively and subsequently considering developments in phase planes of these variables, fast and slow, as if they are independent. When uncoupled, each pair has unique asymptotically stable steady state and no self-sustained oscillatory behavior (although damped oscillations about the equilibrium are admitted). When the consumer-resource pairs are weakly coupled through direct reciprocal inhibition of consumers, the whole system exhibits self-sustained relaxation oscillations with a period that can be significantly longer than intrinsic relaxation time of either pair. It is shown that the model equations adequately describe locally linked consumer-resource systems of quite different nature: living populations under interspecific interference competition and lasers coupled via their cavity losses.
\end{abstract}

\maketitle

\section{Introduction}
Recently, there has been a great deal of activity aimed at studying the synchronization of coupled oscillators of diverse nature \cite{Balanov:2009,Hoppensteadt:1997,Pikovsky:2001,Strogatz:2003,Vandermeer:2006}. The theory of synchronization implies that even in uncoupled state the individual elementary units exhibit self-sustained oscillations. However no less interesting are the systems where local coupling is essential for the very generation of oscillations and not only for their modulation or phase adjustment.

As far back as in early 1970s, Smale \cite{Smale:1974} constructed a counterintuitive mathematical example of a biological cell modeled by the chemical kinetics of four metabolites, $x_{1},\dotsc,x_{4}$, such that the reaction equations $\mathrm{d}\mathbf{x}/\mathrm{d}t = \mathbf{R}(\mathbf{x})$ for the set of metabolites, $\mathbf{x} = (x_{1},\dotsc,x_{4})$, had a globally stable equilibrium. The cell is ``dead'', in that the concentrations of its metabolites always tend to the same fixed levels. When two such cells are coupled by linear diffusion terms of the form $\mathbf{M}(\mathbf{x}_{2} -\mathbf{x}_{1})$, where $\mathbf{M}$ is a diagonal matrix with the elements $\mu_{k}\delta_{kl}$, however, the resulting equations are shown to have a globally stable limit cycle. The concentrations of the metabolites begin to oscillate, and the system becomes ``alive''. In Smale's words:
\begin{quote}
There is a paradoxical aspect to the example. One has two dead (mathematically dead) cells interacting by a diffusion process which has a tendency in itself to equalize the concentrations. Yet in interaction, a state continues to pulse indefinitely.
\end{quote}

The reaction equations involved in Smale's model were too general to appeal to any specific process. Since then, inspired by his pioneer work, a number of models have been proposed containing biologically plausible mechanisms by which coupling of identical nonoscillating cells could generate synchronous oscillations. Among them are a model of electrically coupled cells, characterized by an excitable membrane and calcium dynamics \cite{Loewenstein:2001}, a model in which coupling a passive diffusive cytoplasmic bulk with an excitable membrane (having an activator-inhibitor dynamics) produces a self-sustained oscillatory behavior \cite{Gomez-Marin:2007}, an analog operational-amplifier implementation of neural cells connected by passive coupling (where conductance of the resistive connection simulates the diffusion coefficient) \cite{Szatmari:2008}, etc. The authors of the last model suggested a term 'awakening dynamics' for the phenomenon.

The subject of the present paper is an emergence of collective oscillations in a system of coupled nonoscillatory consumer-resource (CR) pairs. Owing to simplicity, this model in a sense may be considered minimal. Our choice of coupled CR equations as a matter of enquiry is dictated primarily by the ubiquity and importance of CR interactions.

CR models are the fundamental building blocks used in mathematical description and simulation of ecosystems. Depending on a specific nature of the involved CR interactions, they can take the forms of predator-prey, plant-herbivore, parasite-host, and victim-exploiter systems \cite{Murdoch:2003}. However applications of the CR models extend far beyond the ecology and are found wherever one can speak of win-loss interactions. In its broad meaning, resource is any substance which can lead to increased growth rate of the consumer as its availability in the environment is increased. As this takes place, the resource is certainly consumed. Consuming the resource means tending to reduce its availability. When carefully examined, CR models are identified in the following fields: epidemiology (susceptible and infected \cite[ch. 10]{Murray:2002}), laser dynamics (photons and electrons \cite[ch. 6]{Agrawal:1993}), labor economics (share of labor and employment rate \cite[p. 28]{Zhang:1991}), theoretical immunology (antigens and B lymphocytes \cite[p. 299]{Volkenstein:1983}), kinetics of chain chemical reactions (lipid molecules and free radicals \cite{Chernavskii:1977}), and in numerous other studies from diverse disciplines.

We consider the situation when each of two consumer species exploits one respective resource only. As explained in Appendix, terms ``consumer'' and ``resource'' in our model may bear not only their literal ecological meaning, but the physical meaning of photon density and population inversion in a laser cavity as well. Both resources are being supplied with constant rates like in a chemostat and consumed according to a simple mass-action kinetics. The resources are thought to be noninteractive. When uncoupled, self-inhibition of the individual consumer population is due to intraspecific interference. The coupling is assumed to originate solely from the interspecific interference competition between the consumers and quantitatively expressed by a bilinear term combining the competitor densities. Thus, the per individual loss rate of either consumer is proportional to the density of its counterpart.

Representation of competition between species in terms of loss-coupling dates back to the classical model of Lotka--Volterra--Gause (LVG) \cite{Gause:1935}. The LVG model operates with carrying capacities of the species, rather than referring explicitly to any essential resources. As shown by MacArthur \cite{MacArthur:1970}, LVG equations may be considered as a quasi-steady-state approximation to the CR equations accentuating resource-mediated nature of competition, under the assumption of relatively rapid dynamics of the involved resources. Thereafter trophic competition have developed into a major descriptor of competition in the ecological literature generally, and in conceptualizing ecosystems as systems of coupled CR oscillators specifically \cite{Vandermeer:2006}. In contrast to the prevailing models of competition we consider the case of pure interspecific interference competition between the consumers with no consumption-induced contribution. Actually, our model is nothing more nor less than LVG equations augmented with the rate equations for the resources. Another key assumption of the model is that the dynamics of the consumers is much faster than that of the resources.

From a physical perspective, by and large similar equations with the like time hierarchy (fast consumer and slow resource) have been in use for coupled longitudinal modes in a semiconductor laser with an intracavity-doubling crystal since the work of Baer \cite{Baer:1986}. These equations have been treated mostly numerically. The notable analytical result belongs to Erneux and Mandel \cite{Erneux:1995} who succeeded to show that the system admits antiphase periodic solutions by reducing it to the equations for quasi-conservative oscillator. However this result has to do with the onset of low-amplitude quasi-harmonic oscillations. Unlike their study, our approach deals with well-developed high-amplitude essentially nonlinear oscillations. Besides, we propose the model to be valid not only for competing laser modes, but for loss-coupled lasers as well.

We analyze the model using geometric singular perturbation technique according to which the full system of equations is decomposed into fast and slow subsystems. As we shall see below, the model reveals qualitatively different behavior at intense and weak competition between the consumer species. If coupling is strong, one of the consumers wins and completely dominates. When coupling is weak, the model exhibits low-frequency antiphase relaxation oscillations with each species alternatively taking the dominant role.

\section{The model}
The two-consumer, two-resource model we consider is the following nondimensional system of four ordinary differential equations:
\begin{subequations}\label{coupled-uv}
\begin{align}
\dot{u}_{1}& = \gamma_{1} -(u_{1} +1)v_{1} -u_{1},\label{coupled-uv-a}\\
\dot{u}_{2}& = \gamma_{2} -(u_{2} +1)v_{2} -u_{2},\label{coupled-uv-b}\\
\varepsilon \dot{v}_{1}& = (u_{1} -\delta v_{1} -\varkappa_{2}v_{2})v_{1},\label{coupled-uv-c}\\
\varepsilon \dot{v}_{2}& = (u_{2} -\delta v_{2} -\varkappa_{1}v_{1})v_{2}.\label{coupled-uv-d}
\end{align}
\end{subequations}
Here dots indicate differentiation with respect to the nondimensional time variable $t$, $u_{i}$ and $v_{i}$ ($i=1,2$) are quantities measuring the respective population sizes of $i$th resource and $i$th consumer, $\gamma_{i} > 0$ ($i=1,2$) is the inflow rate of $i$th resource, $0 < \delta \ll 1$ is a parameter representing consumer self-limitation, $\varkappa_{j} > 0$ ($j=1,2$ and $j \neq i$) quantifies the inhibitory effect of $j$th consumer on the growth of $i$th consumer due to coupling, and $0 < \varepsilon \ll 1$ is a singular perturbation parameter indicating that the dynamics of the consumers is much faster than that of the resources.

It should be mentioned that being proportional to its dimensional prototype, $v_{i}$ directly represents population density of consumer species and is always nonnegative. Quantity $u_{i}$, however, is not a population size in the true sense of the word. It is rather an affine transformation of a population size of the form $N \to a N + b$, where $a$ and $b$ are scaling constants. This is done for reasons of mathematical convenience. Unlike a purely linear transformation, an affine map does not preserve the zero point, so in \eqref{coupled-uv} $u_{i} = -1$ corresponds to zero population size in reality. Nevertheless, from here on we shall apply the term ``resource'' to $u_{i}$ for brevity.

For more details and discussion on the derivation of the model \eqref{coupled-uv} from different perspectives the reader is referred to Appendix.

\section{Model analysis and implications}
\subsection{A single CR pair} When $\varkappa_{1,2} = 0$, the communities are uncoupled and completely independent. An isolated CR pair is governed by equations
\begin{subequations}\label{isolated-uv}
\begin{align}
\dot{u}& = \gamma -(u +1)v -u,\label{isolated-uv-a}\\
\varepsilon \dot{v}& = (u -\delta v)v.\label{isolated-uv-b}
\end{align}
\end{subequations}

There exist two nonnegative steady states:
\begin{subequations}\label{steady-uv}
\begin{align}
\overline{u}& = \gamma,\quad \overline{v} = 0;\label{steady-uv-a}\\
\overline{u}& = \textstyle{\frac{1}{2}}(\sqrt{1 +(4\gamma +2 +\delta)\delta} -1 -\delta) =\gamma\delta +\mathcal{O}(\delta^{2}),\label{steady-uv-b}\\
\overline{v}& = \textstyle{\frac{1}{2\delta}}(\sqrt{1 +(4\gamma +2 +\delta)\delta} -1 -\delta) =\gamma -\gamma(\gamma +1)\delta +\mathcal{O}(\delta^{2}).\notag
\end{align}
\end{subequations}
The linearization of \eqref{isolated-uv} takes the form
\begin{equation}\label{Jacobian-uv}
\mathbf{J} = \begin{pmatrix}
-\overline{v} -1 & -\overline{u} -1\\
\overline{v}/\varepsilon &(\overline{u} -2\delta\overline{v})/\varepsilon
\end{pmatrix},
\end{equation}
where ($\overline{u}$, $\overline{v}$) is one of the above steady states \eqref{steady-uv}.

\begin{figure}[htbp]
\noindent\centering{
\includegraphics[scale=0.5]{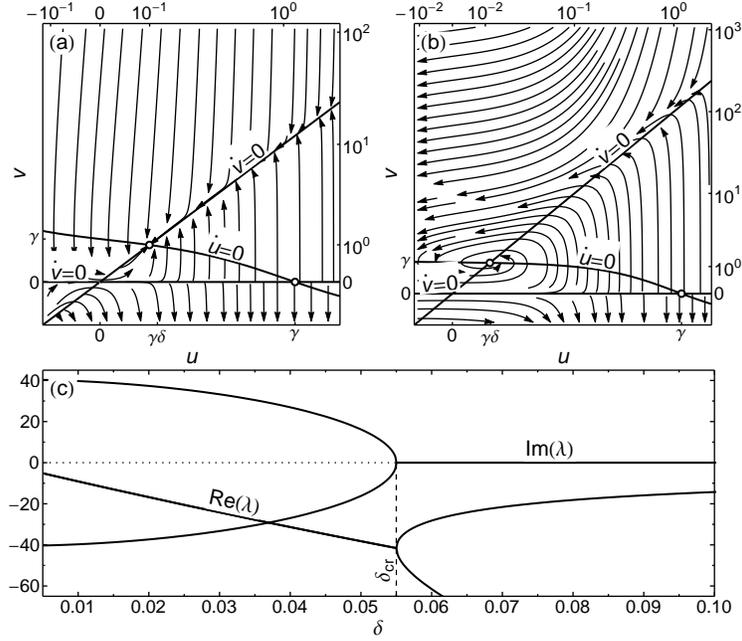}}
\caption{Phase portraits of an isolated CR system \eqref{isolated-uv} for two cases when the unique stable equilibrium \eqref{steady-uv-b} is a node (a), and a focus (b). Shown are vertical ($\dot{u}=0$) and horizontal ($\dot{v}=0$) nullclines. Steady states are marked by open circles. Numerical simulations have been carried out with data from Table \ref{tabA1}. The respective values of the second order loss parameter $\delta$ in (a) and (b) are $0.1$ and $0.01$. Typically, consumer scale ($v$) is enormous; the equilibrium point happens to lie very close to the origin. For better appearance, henceforward we display all the graphs using coordinate transformation $u\to\operatorname{arsinh}(u/\gamma\delta)$ and $v\to\operatorname{arsinh}(v/\gamma)$. (c) Eigenvalues of \eqref{Jacobian-uv} at \eqref{steady-uv-b} as a function of the second order loss $\delta$. Critical value $\delta_{\mathrm{cr}}$ indicates a boundary between the regions of damped oscillations and aperiodic damping.}
\label{fig01}
\end{figure}

At \eqref{steady-uv-a}, one eigenvalue is negative and one is positive: $\lambda_{1}=-1$, $\lambda_{2}=\gamma/\varepsilon$. Thus \eqref{steady-uv-a} is a saddle point. At \eqref{steady-uv-b}, $\operatorname{Tr}\mathbf{J} = -(\delta/\varepsilon +1)\overline{v} -1 <0$ and $\det\mathbf{J} = (1 +2\delta\overline{v} +\delta)\overline{v}/\varepsilon >0$, so the steady state is a stable node/focus. Specifically, focus is the case for
\begin{equation}\label{focus-uv}
(\operatorname{Tr}\mathbf{J})^{2} -4\det\mathbf{J}
=\left(\gamma^2\delta^2 +\mathcal{O}(\delta^3)\right)\varepsilon^{-2} +\left(-4\gamma +\mathcal{O}(\delta)\right)\varepsilon^{-1} +\mathcal{O}(1)<0,
\end{equation}
whence one obtains an asymptotic estimate
\begin{equation}\label{focus-cond-uv}
\delta = o(\varepsilon^{1/2}).
\end{equation}
Fig.~\ref{fig01}c illustrates this focus-node bifurcation numerically. Damped oscillations is a well-known inherent feature of the photon-carrier dynamics in class-B lasers.

An isolated CR system \eqref{isolated-uv} admits therefore, only solution which tends asymptotically towards the unique steady state. Periodic solutions are excluded. However the temporal dynamics of approaching this steady state essentially depends on the interplay between $\varepsilon$ and $\delta$, and is worth another look. There are in fact three timescales, $\mathcal{O}(\varepsilon)$, $\mathcal{O}(1)$ and $\mathcal{O}(\delta)$, involved in the CR system \eqref{isolated-uv} when it is overdamped, and two---$\mathcal{O}(\varepsilon)$ and $\mathcal{O}(1)$---when underdamped.

System \eqref{isolated-uv} is singularly perturbed because the derivative of one of its state variables, $v$, is multiplied by a small positive parameter $\varepsilon$. Singular perturbation cause two-time-scale behavior of the system characterized by the presence of slow and fast transients in the system's response to external stimuli.

Replacing $t$ in \eqref{isolated-uv} with a fast time variable $\tau=t/\varepsilon$ and setting $\varepsilon=0$ we obtain the fast subsystem
\begin{subequations}\label{isolated-fast}
\begin{align}
u'& = 0,\label{isolated-fast-u}\\
v'& = (u -\delta v)v,\label{isolated-fast-v}
\end{align}
\end{subequations}
where prime means differentiation with respect to $\tau$. In the stretched timescale $\tau$ the slow resource variable $u$, according to \eqref{isolated-fast-u}, is replaced by its initial value and reckoned as constant parameter. Equation \eqref{isolated-fast-v} is of a logistic type which has the solution
\begin{equation}\label{logistic}
v(\tau)=\frac{u/\delta}{1 +\left(u/\delta\,v(0) -1\right)\exp(-u\tau)},
\end{equation}
valid for $t = \mathcal{O}(\varepsilon)$.

After a lapse of considerable time (in the fast scale) $v$ converges to either of two fixed points depending on a sign of $u$:
\begin{equation}\label{vatinf}
\lim_{\tau \to \infty} v(\tau) =
\begin{cases}
u/\delta, &\text{if $u > 0$;}\\
0,        &\text{if $u < 0$.}
\end{cases}
\end{equation}
It means that every single trajectory starting within the positive quadrant of $(u,v)$ plane far enough from the stable steady state \eqref{steady-uv-b} will run almost parallel to the vertical axis and hit the line $u -\delta v = 0$ practically in a finite time of order $\mathcal{O}(\varepsilon)$. This is shown in Fig.~\ref{fig01}a--b.

Now set $\varepsilon=0$ in \eqref{isolated-uv} to get the slow subsystem
\begin{subequations}\label{isolated-slow}
\begin{align}
\dot{u}& = \gamma -(u +1)v -u,\label{isolated-slow-u}\\
0& = (u -\delta v)v.\label{isolated-slow-v}
\end{align}
\end{subequations}
The equation \eqref{isolated-slow-v} describes a slow manifold consisting of two lines in the $(u,v)$ plane: $v=u/\delta$ and $v=0$. By \eqref{vatinf}, the former attracts all the trajectories in the first quadrant, while the latter---all those in the second. Inasmuch as the quasi-steady state of \eqref{isolated-fast-v}, $\overline{v}=u/\delta$, is an isolated root of \eqref{isolated-slow-v} and $\overline{v}$ is a stable solution of \eqref{isolated-slow-v} for any $u>0$, the assumptions of Tikhonov's theorem \cite{Tikhonov:1952} are satisfied, and one may proceed to approximate $u$ and $v$ in terms of the solution of the reduced system \eqref{isolated-slow} in the slow timescale. It means that after arriving at the slow manifold $v=u/\delta$, the representing point of the full system \eqref{isolated-uv} will move along the manifold toward the equilibrium point with a characteristic velocity of order $\mathcal{O}(1)$.

In the immediate proximity to equilibrium \eqref{steady-uv-b} the behavior of the trajectory is determined by the type of the fixed point, whether it is a node or a focus. Eventually this depends on the value of $\delta$. Namely, close to a stable node, the system has two distinct real negative eigenvalues, one fast ($\lambda_{1}$), and one slow ($\lambda_{2}$):
\begin{equation}\label{eigenvalue-node}
\begin{aligned}
\lambda_{1}&= \left(-\gamma\delta +\mathcal{O}(\delta^{2})\right)\varepsilon^{-1} +\left(\delta^{-1} +\mathcal{O}(1)\right) +\mathcal{O}(\varepsilon),\\
\lambda_{2}&= \left(-\delta^{-1} +\mathcal{O}(1)\right) +\mathcal{O}(\varepsilon),
\end{aligned}
\qquad \text{for } 1>\delta\gg\varepsilon.
\end{equation}
Since $\left|\lambda_{1}\right| \gg \left|\lambda_{2}\right|$, trajectories starting off the associated eigenvector 2 (which is tangent to the slow manifold $v=u/\delta$) converge to that vector along lines almost parallel to eigenvector 1 (which is parallel to the vertical axis $v$). As they approach vector 2 they become tangent to it and move along it up to the very nodal point (Fig.~\ref{fig01}a). The characteristic time constant of this final stage is of order $\mathcal{O}(\delta)$.

In the vicinity of a stable focus the motion is qualitatively different: trajectories still keep converging to the equilibrium, but no longer follow the slow manifold $v=u/\delta$ (Fig.~\ref{fig01}b). The reason is that for a focus eigenvectors are complex. Recalling that condition \eqref{focus-cond-uv} must be true for a focus, calculate the eigenvalues in a limit case of $\delta \to 0$:
\begin{equation}\label{eigenvalue-focus}
\lambda_{1,2} = -\textstyle\frac{1}{2}(\gamma +1) \pm\mathrm{i}\sqrt{{\gamma}/{\varepsilon}}\qquad \text{for } \delta=0.
\end{equation}
Loosely speaking, one may think that near a focal point separation of state variables into slow and fast ones ceases to have its conventional meaning. There is no point to talk about motion along the slow manifold since there are no reduced one-dimensional systems corresponding to the neighborhood of a focus. The perturbation parameter $\varepsilon$ affects only the frequency of damped oscillation, but not the damping rate. The radius of the focal spiral uniformly shrinks with a time constant of order $\mathcal{O}(1)$.

To $\mathcal{O}(1)$ for small $\delta$ there is a way to recast the equations \eqref{isolated-uv} near the focal equilibrium $(0,\gamma)$ in a convenient form where the perturbation parameter $\varepsilon$ does not multiply any right hand side. Namely, performing the scaling $t=\mu^{2}s$, $u=\mu^{2} \gamma \xi$ and $v=\gamma (1+\eta)$, where $\mu^{2}=\sqrt{\varepsilon/\gamma}$, one obtains
\begin{equation}\label{hamilton}
\begin{split}
\dot{\xi}& = -\eta -\mu^{2}\xi\bigl(1+\gamma(1+\eta)\bigr),\\
\dot{\eta}& = \xi(1+\eta).
\end{split}
\end{equation}
Here dots stand for differentiation with respect to the time variable $s$.

\begin{figure}[htbp]
\noindent\centering{
\includegraphics[scale=1]{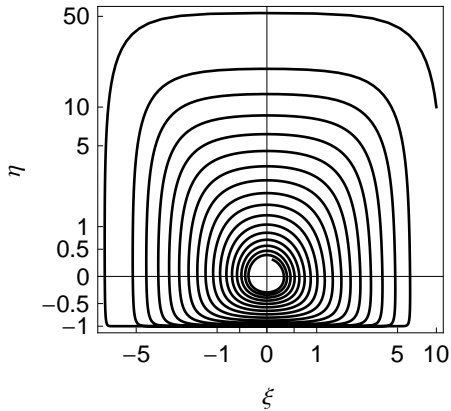}}
\caption{Phase-plane trajectory of the weakly perturbed Hamiltonian system \eqref{hamilton} for $\gamma = 1.19375$, $\varepsilon = 0.727273\times 10^{-3}$, and $\mu = 0.157107$. Near the equilibrium the oscillations are almost harmonic.}
\label{fig02}
\end{figure}

Equations \eqref{hamilton} represent a weakly perturbed Hamiltonian system. For $\mu=0$ the system is pure Hamiltonian and admits the first integral
\[
H=\textstyle\frac{1}{2}\xi^{2}+\eta-\ln(1+\eta),
\]
which is a conserved quantity ($\dot{H}=0$). The periodic solutions of the Hamiltonian system form a one-parameter family with the equilibrium $(0,0)$ as center point. The condition $0< \mu \ll 1$ makes equations \eqref{hamilton} quasi-conservative with phase trajectories slowly spiralling to the equilibrium (Fig.~\ref{fig02}).

System \eqref{hamilton} can be rewritten as a second order differential equation for $\eta$ only:
\begin{equation}\label{quasicons1}
\ddot{\eta} +\eta +\eta^{2} -\frac{\dot{\eta}^{2}}{1 +\eta} +\mu^{2}\bigl(1 +\gamma(1 +\eta)\bigr)\dot{\eta} = 0.
\end{equation}
This is the equation for a nonlinear quasi-conservative oscillator. As is seen from \eqref{quasicons1}, for $|\eta|\ll 1$ and $|\dot{\eta}|\ll 1$ the oscillator is not only quasi-conservative, but also a quasi-linear with the frequency $\omega_{0}=1$. These conditions motivate introducing the new variable $\zeta = \eta/\mu$. As a result one obtains
\begin{equation}\label{quasicons2}
\ddot{\zeta} +\zeta = \mu\Bigl(-\zeta ^2 +\frac{{\dot{\zeta}}^{2}}{1 +\mu \zeta} -\mu(1 +\gamma + \mu \gamma \zeta)\dot{\zeta}\Bigr).
\end{equation}
This equation of quasi-conservative, quasi-linear oscillator will be used later on when analyzing the onset of synchronous periodicity.

\subsection{Coupled communities: Stability}
In the following, we will distinguish two main types of coupling: strong, $\varkappa_{1,2}>1$, and weak, $\varkappa_{1,2}<1$.

Physically feasible equilibria, or fixed points, $(\overline{u}_{1},\overline{u}_{2},\overline{v}_{1},\overline{v}_{2})$, of \eqref{coupled-uv} are those for which $\overline{v}_{1,2}\geqslant 0$. We denote the interior fixed point by $F_{12}=(\overline{u}_{1}, \overline{u}_{2},\overline{v}_{1},\overline{v}_{2})$, where the subscripts stand for the consumers. Lack of a certain index at a boundary fixed point means that the consumer concerned is not present (extinct). Thus $F_{1}=(\overline{u}_{1},\overline{u}_{2},\overline{v}_{1},0)$ and $F_{2}=(\overline{u}_{1},\overline{u}_{2},0,\overline{v}_{2})$ designate either of one-consumer equilibria corresponding to dominance, while $F=(\overline{u}_{1},\overline{u}_{2},0,0)$ means both consumers having been washed out.

Model \eqref{coupled-uv} has four feasible steady states. To $\mathcal{O}(1)$ for small $\delta$
\begin{subequations}\label{ss}
\begin{alignat}{3}
F:\qquad&\overline{u}_{1}=\gamma_{1},\quad \overline{u}_{2}=\gamma_{2},\quad \overline{v}_{1}=0,\quad \overline{v}_{2}=0;\label{ssF}\\
F_{1}:\qquad&\overline{u}_{1}=0,\quad \overline{u}_{2}=\gamma_{2},\quad \overline{v}_{1}=\gamma_{1},\quad \overline{v}_{2}=0;\label{ssF1}\\
F_{2}:\qquad&\overline{u}_{1}=\gamma_{1},\quad \overline{u}_{2}=0,\quad \overline{v}_{1}=0,\quad \overline{v}_{2}=\gamma_{2};\label{ssF2}\\
F_{12}:\qquad&\overline{u}_{1}=\frac{\varkappa_{1}\gamma_{1} -\varkappa_{2}\gamma_{2} -\varkappa_{1}\varkappa_{2} +1 \pm R}{2(\varkappa_{1} -1)},\notag\\
&\overline{u}_{2}=\frac{-\varkappa_{1}\gamma_{1} +\varkappa_{2}\gamma_{2} -\varkappa_{1}\varkappa_{2} +1 \pm R}{2(\varkappa_{2} -1)},\notag\\
&\overline{v}_{1}={\overline{u}_{2}}/{\varkappa_{1}},\quad \overline{v}_{2}={\overline{u}_{1}}/{\varkappa_{2}};\label{ssF12}
\end{alignat}
\end{subequations}
where $R=\sqrt{(\varkappa_{1}\gamma_{1} -\varkappa_{2}\gamma_{2} -\varkappa_{1}\varkappa_{2} +1)^{2} +4\varkappa_{2}(\varkappa_{1} -1)(\varkappa_{1}\gamma_{1} -\gamma_{2})}$.

Boundary equilibria $F$, $F_{1}$ and $F_{2}$ always exist. $F_{12}$ exists if
\begin{subequations}\label{existF12}
\begin{alignat}{5}
1/\varkappa_{2}&<\gamma_{2}/\gamma_{1}<\varkappa_{1}\quad && \mathrm{for}\quad \varkappa_{1}>1 \wedge \varkappa_{2}>1\quad \mathrm{(strong\ coupling)},\label{existF12strong}\\
\intertext{or}
\varkappa_{1}&<\gamma_{2}/\gamma_{1}<1/\varkappa_{2}\quad && \mathrm{for}\quad \varkappa_{1}<1 \wedge \varkappa_{2}<1\quad \mathrm{(weak\ coupling)}\label{existF12weak}.
\end{alignat}
\end{subequations}
In case \eqref{existF12strong} the expression with ``$+$'' in \eqref{ssF12} is realized, while in case \eqref{existF12weak} the one with ``$-$'' is feasible.

$F$ is always unstable, because two of the four associated eigenvalues are positive:
\begin{equation}\label{lambdaF}
\lambda(F):\qquad \gamma_{1}/\varepsilon,\quad \gamma_{2}/\varepsilon,\quad -1,\quad -1.
\end{equation}

Correct to $\mathcal{O}(1)$ in $\varepsilon$, the eigenvalues for $F_{1}$ and $F_{2}$ are
\begin{subequations}\label{lambdaF1F2}
\begin{align}
\lambda(F_{1}):\qquad& -1,\quad -\textstyle\frac{1}{2}(\gamma_{1} +1) \pm\mathrm{i}\sqrt{{\gamma_{1}}/{\varepsilon}},\quad
({\gamma_{2} -\varkappa_{1}\gamma_{1}})/{\varepsilon};\\
\lambda(F_{2}):\qquad& -1,\quad -\textstyle\frac{1}{2}(\gamma_{2} +1) \pm\mathrm{i}\sqrt{{\gamma_{2}}/{\varepsilon}},\quad
({\gamma_{1} -\varkappa_{2}\gamma_{2}})/{\varepsilon},
\end{align}
\end{subequations}
whence it follows that \eqref{ssF1} and \eqref{ssF2} are stable when
\begin{subequations}\label{stabilityF1F2}
\begin{align}
\gamma_{2}/\gamma_{1}&<\varkappa_{1}\label{stabilityF1}\\
\intertext{and}
\gamma_{1}/\gamma_{2}&<\varkappa_{2},\label{stabilityF2}
\end{align}
\end{subequations}
respectively.

The necessary and sufficient conditions for all the eigenvalues of the Jacobian matrix,
\begin{equation}\label{JacobianF12}
\begin{pmatrix}
-\overline{v}_{1} -1 & 0 & -\varkappa_{2}\overline{v}_{2} -1 & 0\\
0 & -\overline{v}_{2} -1 & 0 & -\varkappa_{1}\overline{v}_{1} -1\\
\overline{v}_{1}/\varepsilon & 0 & 0 & -\varkappa_{2}\overline{v}_{1}/\varepsilon\\
0 & \overline{v}_{2}/\varepsilon & -\varkappa_{1}\overline{v}_{2}/\varepsilon & 0
\end{pmatrix},
\end{equation}
evaluated at $F_{12}$, to have negative real parts are, from the Routh--Hurwitz criterion,
\begin{subequations}\label{Routh}
\begin{align}
c_{0}& >0,\label{Routh-a}\\
c_{3}& >0,\label{Routh-b}\\
c_{2}c_{3} -c_{1}&>0,\label{Routh-c}\\
c_{1}(c_{2}c_{3} -c_{1}) -c_{0}c_{3}^{2}& >0,\label{Routh-d}
\end{align}
\end{subequations}
where $c_{0}$, $c_{1}$, $c_{2}$ and $c_{3}$ are the coefficients of the characteristic polynomial $\lambda^{4} +c_{3}\lambda^{3} +c_{2}\lambda^{2} +c_{1}\lambda +c_{0}$ of \eqref{JacobianF12}:
\begin{equation}\label{Routh-coef}
\begin{split}
c_{0}& =\overline{v}_{1}\overline{v}_{2}(\varkappa_{1}(\overline{v}_{1} -\varkappa_{2}(\overline{v}_{1} +\overline{v}_{2} +1)) +\varkappa_{2}\overline{v}_{2} +1)\varepsilon^{-2},\\
c_{1}& =-\varkappa_{1}\varkappa_{2}\overline{v}_{1}\overline{v}_{2}(\overline{v}_{1} +\overline{v}_{2} +2)\varepsilon^{-2}\\
&\quad +(\overline{v}_{1}(\overline{v}_{2}(\varkappa_{1} +\varkappa_{2} +\varkappa_{1}\overline{v}_{1} +\varkappa_{2}\overline{v}_{2} +2) +1) +\overline{v}_{2})\varepsilon^{-1},\\
c_{2}& =-\varkappa_{1}\varkappa_{2}\overline{v}_{1}\overline{v}_{2}\varepsilon^{-2} +(\overline{v}_{1}(\overline{v}_{2}(\varkappa_{1}+\varkappa_{2}) +1) +\overline{v}_{2})\varepsilon^{-1} +(\overline{v}_{1} +1)(\overline{v}_{2} +1),\\
c_{3}& =\overline{v}_{1} +\overline{v}_{2} +2.
\end{split}
\end{equation}

To analyze the validity of \eqref{Routh} we put $\varkappa_{2}=\mu\varkappa_{1}$, where $\mu=\mathcal{O}(1)$. Then for \eqref{Routh-a} to be true, $\varkappa_{1}=o(1)$ should be met, which is incompatible with strong coupling, yet possible for weak coupling. Conditions \eqref{Routh-b} and \eqref{Routh-c} are always valid. As is known, \eqref{Routh-d} guarantees a simple complex conjugate pair of eigenvalues corresponding to a linearization about steady state $F_{12}$ to have negative real part. For $\varkappa_{1}=o(1)$ it boils down to
\begin{multline}\label{Routh-d-reduced}
\left(-\overline{v}_{1}\overline{v}_{2}(\overline{v}_{1} +\overline{v}_{2} +2)(\overline{v}_{1} +\overline{v}_{2} +\overline{v}_{1}^{2} +\overline{v}_{2}^{2})\mu\varkappa_{1}^{2} +\mathcal{O}(\varkappa_{1}^{3})\right)\varepsilon^{-3}\\
 +\left((\overline{v}_{1} +1)(\overline{v}_{2} +1)(\overline{v}_{1} -\overline{v}_{2})^{2} +\mathcal{O}(\varkappa_{1})\right)\varepsilon^{-2} +\mathcal{O}(\varepsilon^{-1})>0,
\end{multline}
whence it follows that
\begin{equation}\label{stability}
\varkappa_{1}=o(\varepsilon^{1/2}).
\end{equation}
With regard to a fairly small value of $\varepsilon$, \eqref{stability} may be thought to be broken under most physically meaningful conditions unless coupling is infinitesimally weak. Hence, normally, condition \eqref{Routh-d} of the Routh--Hurwitz criterion is never fulfilled and the interior fixed point $F_{12}$---if it exists---is always unstable by growing oscillations.

Existence and stability conditions of possible nonnegative equilibrium points are summarized in Table \ref{tab1}.
\begin{table}[htbp]
\caption{Existence and stability conditions of nonnegative equilibria of system \eqref{coupled-uv}.}
\begin{center}\footnotesize
\begin{tabular}{lll}
\toprule
Equilibrium & Existence & Stability \\
\midrule
$F$       & Always & Unstable \\
$F_{1}$   & Always & $\gamma_{2}/\gamma_{1}<\varkappa_{1}$ \\
$F_{2}$   & Always & $\gamma_{2}/\gamma_{1}>1/\varkappa_{2}$ \\
$F_{12}$  & $1/\varkappa_{2}<\gamma_{2}/\gamma_{1}<\varkappa_{1}$\quad for\quad $\varkappa_{1,2}>1$\ (strong coupling) &Unstable\\
          & $\varkappa_{1}<\gamma_{2}/\gamma_{1}<1/\varkappa_{2}$\quad for\quad $\varkappa_{1,2}<1$ (weak coupling) &$\varkappa_{1,2}=o(\varepsilon^{1/2})$\\
\bottomrule
\end{tabular}
\end{center}
\label{tab1}
\end{table}

\subsection{Strong coupling: Bistability and hysteresis}
\begin{figure}[htbp]
\noindent\centering{
\includegraphics[scale=0.6]{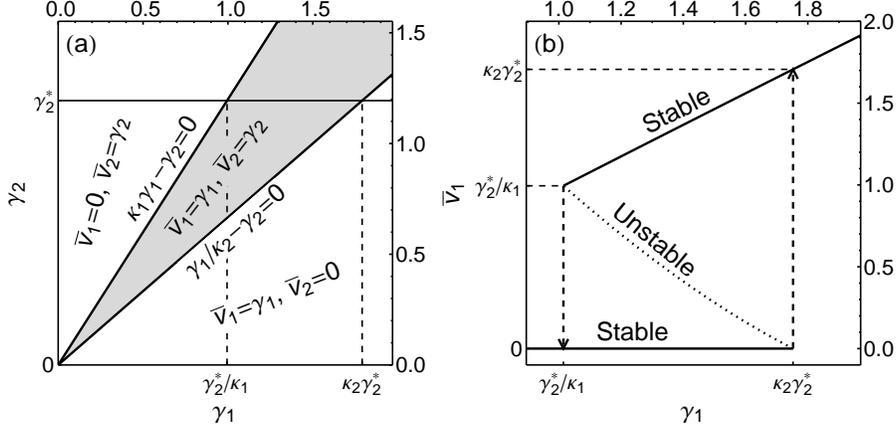}}
\caption{Hysteresis in the two-consumer, two-resource system \eqref{coupled-uv}. It takes place in the case of strong coupling $\varkappa_{1,2}>1$. (a) The resource-supply parametric plane $(\gamma_{1},\gamma_{2})$. Consumer 1 completely dominates below the line $\gamma_{1} -\varkappa_{2}\gamma_{2} =0$, whereas consumer 2---above the line $\varkappa_{1}\gamma_{1} -\gamma_{2} =0$. Both boundary steady states are stable in the region of bistability confined by the two aforementioned lines and marked off by gray, with the dominance of either consumer being a matter of path-dependency. (b) Equilibrium level of consumer 1 as a function of the resource supply. Numerical values of the coupling strengths are chosen to be $\varkappa_{1}=1.2$ and $\varkappa_{2}=1.5$.}
\label{fig03}
\end{figure}

Strong coupling in \eqref{coupled-uv} makes possible bistability of boundary equilibria, as evident from Table \ref{tab1}. When both $F_{1}$ and $F_{2}$ are stable with an unstable coexistence steady state $F_{12}$, the system being studied is able to exhibit a hysteresis effect. Given the strong coupling, suppose, that the inflow of resource 2 is kept at some constant level, $\gamma_{2} = \gamma_{2}^{\ast}$, while the inflow of resource 1, $\gamma_{1}$, steadily increases from a value less than $\gamma_{2}^{\ast}/\varkappa_{1}$ along the line $\gamma_{2} = \gamma_{2}^{\ast}$ in the $(\gamma_{1}, \gamma_{2})$ parameter plane, as shown in Fig.~\ref{fig03}a. Referring to \eqref{ss} and \eqref{stabilityF1F2}, one sees that initially $F_{1}$ is unstable, $F_{2}$ is stable and $F_{12}$ does not exist with the complete dominance of consumer 2. Within the interval $\gamma_{2}^{\ast}/\varkappa_{1}<\gamma_{1}<\varkappa_{2}\gamma_{2}^{\ast}$ steady state $F_{1}$ becomes stable yet empty and $F_{12}$ becomes existent (according to \eqref{existF12strong}) yet unstable, so the situation remains unchanged until point $(\varkappa_{2}\gamma_{2}^{\ast},\gamma_{2}^{\ast})$ in Fig.~\ref{fig03}a has been reached from the left. For a larger $\gamma_{1}$, state $F_{2}$ gives up its stability and the system jumps to $F_{1}$. Consumer 2 gets washed out, while consumer 1 takes over. If now we start reducing $\gamma_{1}$, the system remains in $F_{1}$ until $\gamma_{1}$ drops to the lower critical value $\gamma_{2}^{\ast}/\varkappa_{1}$, beyond which $F_{1}$ is no longer stable and there is a reverse jump to $F_{2}$. In other words, as $\gamma_{1}$ progresses along $\gamma_{2} = \gamma_{2}^{\ast}$ there is a discontinuous switch from consumer 2 to consumer 1 at $\varkappa_{2}\gamma_{2}^{\ast}$, while as $\gamma_{1}$ retraces its steps, there is a discontinuous switch from consumer 1 to consumer 2 at $\gamma_{2}^{\ast}/\varkappa_{1}$. Fig.~\ref{fig03}b illustrates how the steady state level of consumer 1 responds to infinitesimally slow changes of the inflow rate of its own resource. The hysteresis is made possible thanks to the concurrent stability of both boundary equilibria and instability of the interior fixed point for $\gamma_{1}\in(\gamma_{2}^{\ast}/\varkappa_{1},\varkappa_{2}\gamma_{2}^{\ast})$. In terms of electronics, such a situation would describe a flip-flop circuit---bistable multivibrator---having two stable conditions, each corresponding to one of two alternative input signals.

\subsection{Weak coupling: Antiphase relaxation oscillations}
As seen from Table \ref{tab1}, the very existence of interior equilibrium $F_{12}$ in a case of weak coupling (condition \eqref{existF12weak}) implies instability of both boundary fixed points, $F_{1}$ and $F_{2}$. System \eqref{coupled-uv} happens to possess four nonnegative steady states, none of them being stable. As we have found, $F_{12}$ is unstable through growing oscillations. In such a case, the model would thus be expected to have a limit cycle in its four-dimensional phase space corresponding to self-sustained oscillations.

As is known, Hopf bifurcations come in both super- and subcritical types. If a small, attracting limit cycle appears immediately after the fixed point goes unstable, and if its amplitude shrinks back to zero as the parameter is reversed, the bifurcation is supercritical; otherwise, it's probably subcritical, in which case the nearest attractor might be far from the fixed point, and the system may exhibit hysteresis as the parameter is reversed.

To check whether the bifurcation is supercritical or subcritical, we employ the averaging method \cite{Sanders:2007}. For negligible $\delta$ the problem reduces to weakly coupled quasi-conservative oscillators. Introduce the new time $s=t/\mu^{2}$, where $\mu^{2}=\sqrt{\varepsilon/\gamma_{1}}$. The new dynamic variables $\xi_{1,2}$ and $\eta_{1,2}$ are defined by formulas $\xi_{1,2}=u_{1,2}/\mu^{2}\gamma_{1,2}$ and $\eta_{1,2}=(v_{1,2}-\gamma_{1,2})/\gamma_{1,2}$. Thus the variables $\xi_{1,2}$ and $\eta_{1,2}$ are the respective deviations of $u_{1,2}$ and $v_{1,2}$ from their standalone steady state values. With these new variables equations \eqref{coupled-uv} become
\begin{subequations}\label{hamilton4}
\begin{align}
\dot{\xi}_{1}& = -\eta_{1} -\mu^{2}\xi_{1}\bigl(1+\gamma_{1}(1+\eta_{1})\bigr),\label{hamilton4-1}\\
\dot{\eta}_{1}& = \xi_{1}(1+\eta_{1}) -(\gamma_{2}/\gamma_{1})\rho_{2}(1+\eta_{1})(1+\eta_{2}),\label{hamilton4-2}\\
\dot{\xi}_{2}& = -\eta_{2} -\mu^{2}\xi_{2}\bigl(1+\gamma_{2}(1+\eta_{2})\bigr),\label{hamilton4-3}\\
\dot{\eta}_{2}& = (\gamma_{2}/\gamma_{1})\xi_{2}(1+\eta_{2}) -\rho_{1}(1+\eta_{1})(1+\eta_{2}),\label{hamilton4-4}
\end{align}
\end{subequations}
where dots mean differentiation with respect to $s$ and $\rho_{1,2}=\varkappa_{1,2}/\mu^{2}$ are the scaled coupling strengths.

For brevity, we restrict our consideration to the case of identical CR pairs and symmetric coupling setting $\gamma_{1,2}=\gamma$ and $\rho_{1,2}=\rho$. With these assumptions system \eqref{hamilton4} can be transformed to two coupled second-order equations for quasi-conservative, quasi-linear oscillators. This can be done as follows. First, \eqref{hamilton4-2} and \eqref{hamilton4-4} are solved for $\xi_{1}$ and $\xi_{2}$. Secondly, \eqref{hamilton4-2} and \eqref{hamilton4-4} are differentiated with respect to time to get $\ddot{\eta}_{1}$ and $\ddot{\eta}_{2}$. Thirdly, $\xi_{1}$, $\xi_{2}$, $\dot{\xi}_{1}$ and $\dot{\xi}_{2}$ are plugged into the equations for $\ddot{\eta}_{1}$ and $\ddot{\eta}_{2}$. And finally, $\eta_{1,2}$ are replaced by the new variables $\zeta_{1,2}$: $\eta_{1,2}=\mu\zeta_{1,2}$.

As a consequence, we arrive at the following coupled equations:
\begin{equation}\label{zetaeq}
\ddot{\zeta}_{1,2} + \zeta_{1,2} = -\rho \dot{\zeta}_{2,1} +\mu Z_{1}^{(1,2)} +\mu^{2}Z_{2}^{(1,2)} +\mathcal{O}(\mu^{3}),
\end{equation}
where
\begin{equation}\notag
\begin{split}
Z_{1}^{(1,2)}&= -\rho(\gamma +1) -\zeta_{1,2}^{2} +\dot{\zeta}_{1,2}^{2} -\rho\zeta_{1,2}\dot{\zeta}_{2,1},\\
Z_{2}^{(1,2)}&= -(\gamma +1) \bigl(\dot{\zeta}_{1,2} +\rho\zeta_{2,1}\bigr) -\zeta_{1}\bigl(\rho(2\gamma +1) +\dot{\zeta}_{1,2}^{2}\bigr).
\end{split}
\end{equation}

We seek the solution of \eqref{zetaeq} in the quasi-harmonic form
\begin{equation}\notag
\begin{split}
\zeta_{1,2}&=a_{1,2}\cos(\omega s +\varphi_{1,2}),\\
\dot{\zeta}_{1,2}&=-a_{1,2}\omega\sin(\omega s +\varphi_{1,2}),
\end{split}
\end{equation}
where $a_{1,2}$ and $\varphi_{1,2}$ are slowly varying amplitudes and phases, and $\omega$ is the frequency of the synchronous oscillations. Differentiating the assumed form of $\zeta_{1,2}$ and equating the result to the assumed form of $\dot{\zeta}_{1,2}$ yields the first pair of relationships between $a_{1,2}$ and $\varphi_{1,2}$:
\begin{equation}\label{kb-relationship}
\dot{a}_{1,2}\cos(\omega s +\varphi_{1,2}) -a_{1,2}\dot{\varphi}_{1,2}\sin(\omega s +\varphi_{1,2}) = 0.
\end{equation}

Then, differentiating $\dot{\zeta}_{1,2}$ and substituting the resulting expression for $\ddot{\zeta}_{1,2}$ as well as the assumed forms for $\zeta_{1,2}$ and $\dot{\zeta}_{1,2}$ into \eqref{zetaeq} yields the second pair of equations relating $a_{1,2}$ and $\varphi_{1,2}$. Separating into equations for the rate of change of $a_{1,2}$ and $\varphi_{1,2}$ one obtains
\begin{equation}\label{a-phi-exact}
\begin{split}
\dot{a}_{1,2}&= \omega^{-1}\sin(\omega s +\varphi_{1,2})\bigl(A_{0}^{(1,2)} +\mu A_{1}^{(1,2)} +\mu^{2} A_{2}^{(1,2)}\bigr),\\
\dot{\varphi}_{1,2}&= a_{1,2}^{-1}\omega^{-1}\cos(\omega s +\varphi_{1,2})\bigl(A_{0}^{(1,2)} +\mu A_{1}^{(1,2)} +\mu^{2} A_{2}^{(1,2)}\bigr),
\end{split}
\end{equation}
where
\begin{equation}\label{As}
\begin{split}
A_{0}^{(1,2)} &= -a_{1,2}(\omega^2 -1)\cos(\omega s +\varphi_{1,2}) -\rho a_{2,1}\omega\sin(\omega s +\varphi_{2,1}),\\
A_{1}^{(1,2)} &= a_{1,2}\bigl(-a_{1,2}\omega^2\sin^2(\omega s +\varphi_{1,2}) +a_{1,2}\cos^2(\omega s +\varphi_{1,2})\bigr.\\
      &\bigl.\phantom{=\ }{}-\rho a_{2,1}\omega\cos(\omega s +\varphi_{1,2})\sin(\omega s +\varphi_{2,1})\bigr) +\rho(\gamma +1),\\
A_{2}^{(1,2)} &= a_{1,2}\cos(\omega s +\varphi_{1,2})\bigl(a_{1,2}^2\omega^2\sin^{2}(\omega s +\varphi_{1,2}) +\rho(2\gamma +1)\bigr)\\
      &\phantom{=\ }{}+(\gamma +1)\bigl(\rho a_{2,1}\cos(\omega s +\varphi_{2,1}) -a_{1,2}\omega\sin(\omega s +\varphi_{1,2})\bigr).
\end{split}
\end{equation}

To this point no approximations have been made except of the expansion in powers of $\mu$ in \eqref{zetaeq}. Averaging equations \eqref{a-phi-exact} over the period $2\pi/\omega$ and considering $a_{1,2}$, $\varphi_{1,2}$, $\dot{a}_{1,2}$ and $\dot{\varphi}_{1,2}$ to be constants while performing the averaging, one obtains the following equations describing the slow variations of $a_{1,2}$ and $\varphi_{1,2}$:
\begin{subequations}\label{aver1}
\begin{align}
\dot{a}_{1,2}&= \frac{1}{2\pi}\int_{0}^{2\pi/\omega}\bigl(A_{0}^{(1,2)} +\mu A_{1}^{(1,2)} +\mu^{2} A_{2}^{(1,2)}\bigr)\sin(\omega s +\varphi_{1,2})\,\mathrm{d}s\notag\\
&={}-\frac{\rho a_{2,1}\omega\cos\phi +\mu^{2}(\gamma +1)\bigl(a_{1,2}\omega -\rho a_{2,1}\sin\phi\bigr)}{2\omega},\label{aver1-1}\\
\dot{\varphi}_{1,2}&= \frac{1}{2\pi a_{1,2}}\int_{0}^{2\pi/\omega}\bigl(A_{0}^{(1,2)} +\mu A_{1}^{(1,2)} +\mu^{2} A_{2}^{(1,2)}\bigr)\cos(\omega s +\varphi_{1,2})\,\mathrm{d}s\notag\\
&=\frac{\rho(a_{1,2}^{2} +a_{2,1}^{2})\sin\phi}{2a_{1,2}a_{2,1}} +\mathcal{O}(\mu^{2}),\label{aver1-2}
\end{align}
\end{subequations}
where $\phi=\varphi_{1,2}-\varphi_{2,1}$.

Any stable fixed points of \eqref{aver1} could mean that the phase difference between the coupled oscillators do not change in time ($\phi=\mathrm{const}$), and the oscillations are periodic with constant amplitudes $a_{1,2}$. Thus, finding the conditions when these fixed points are stable would mean finding the conditions at which the synchronization occurs. Equations \eqref{aver1} show that the first approximation of the averaging method does not predict any nonzero steady-state amplitudes. The reason for this seems to lie in the fact that the even terms in \eqref{As} do not contribute to the value of integral \eqref{aver1-1}. To take account of them we have to employ in what follows the second approximation of the averaging method. As to the phase difference, $\phi$, we see that stable is only antiphase steady-state regime with $\phi=\pi$.

Let $\mathbf{R}$ be the vector of the right-hand sides of \eqref{a-phi-exact}. Also denote the operation of averaging by the angle brackets $\langle\cdot \rangle$. Then $\langle\mathbf{R} \rangle$ will mean the right-hand sides of \eqref{aver1}. Following the conventional procedure we retain only vibrational terms in $\mathbf{R}$,
\[
\widetilde{\mathbf{R}}=\mathbf{R}-\langle\mathbf{R} \rangle,
\]
and integrate $\widetilde{\mathbf{R}}$ up to an arbitrary function of the amplitudes chosen for simplicity to be equal to zero:
\[
\widehat{\mathbf{R}} = \int \widetilde{\mathbf{R}}\,\mathrm{d}s.
\]
Now, upon calculation of the Jacobian matrix ${\partial \mathbf{R}}/{\partial (a_{1},a_{2},\varphi_{1},\varphi_{2})^{T}}$, we can write down the second approximation symbolically as
\[
(\dot{a}_{1},\dot{a}_{2},\dot{\varphi}_{1},\dot{\varphi}_{2})^{T} = \langle\mathbf{R} \rangle +\Bigl\langle \frac{\partial \mathbf{R}}{\partial (a_{1},a_{2},\varphi_{1},\varphi_{2})^{T}}\,\widehat{\mathbf{R}} \Bigr\rangle.
\]
(The terms now neglected by averaging are of a higher order of magnitude with respect to the small parameters than the terms neglected in the first approximation.)

Due to the symmetry of the case we may set $a_{1}=a_{2}=a$ enabling us to present the resulting equations of the second approximation as compact as possible:
\begin{subequations}\label{aver2}
\begin{align}
\dot{a}&=-\frac{1}{2} a\rho \cos\phi +\frac{\mu^{2}}{24\omega^{3}}a(\rho(\omega\cos\phi(a^2 (2 \omega^2 +5) +12\rho(\gamma +1))\notag\\
&\phantom{=\ }{}+2\omega^2\sin\phi(3a^2 \rho \cos\phi -\rho a^2 +6(\gamma +1)) -2 a^2\omega(\omega^2 +1) \cos 2\phi\notag\\
&\phantom{=\ }{}-6(\gamma +1) \sin\phi) -12(\gamma +1) \omega^3),\label{aver2-1}\\
\dot{\phi}&=\rho\sin\phi +\frac{\mu^{2}}{24\omega^{2}}\rho\sin\phi(3a^2(3\omega^2 -10) +8a^2(\omega^2 +1)\cos\phi\notag\\
&\phantom{=\ }{}-24\rho(\gamma +1)).\label{aver2-2}
\end{align}
\end{subequations}

The stable steady-state solution for the phase difference is $\phi=\pi$. Inserting this value into \eqref{aver2-1} and putting $\dot{a}=0$ we get three different steady-state solutions for $a$ (including the trivial one), of which only one,
\begin{equation}\label{stable-ampl}
a = \frac{2\sqrt{3}\sqrt{\rho\omega^2 -\mu ^2(\gamma +1)(\rho^2 +\omega^2)}}{\mu\sqrt{\rho(4\omega^2 +7)}},
\end{equation}
being stable. Formula \eqref{stable-ampl} indicates a soft transition to sustained oscillations as the coupling parameter $\rho$ is progressively increased from a critical value. Thus we expect that the Hopf bifurcation is supercritical.

\begin{figure}[htbp]
\noindent\centering{
\includegraphics[scale=0.95]{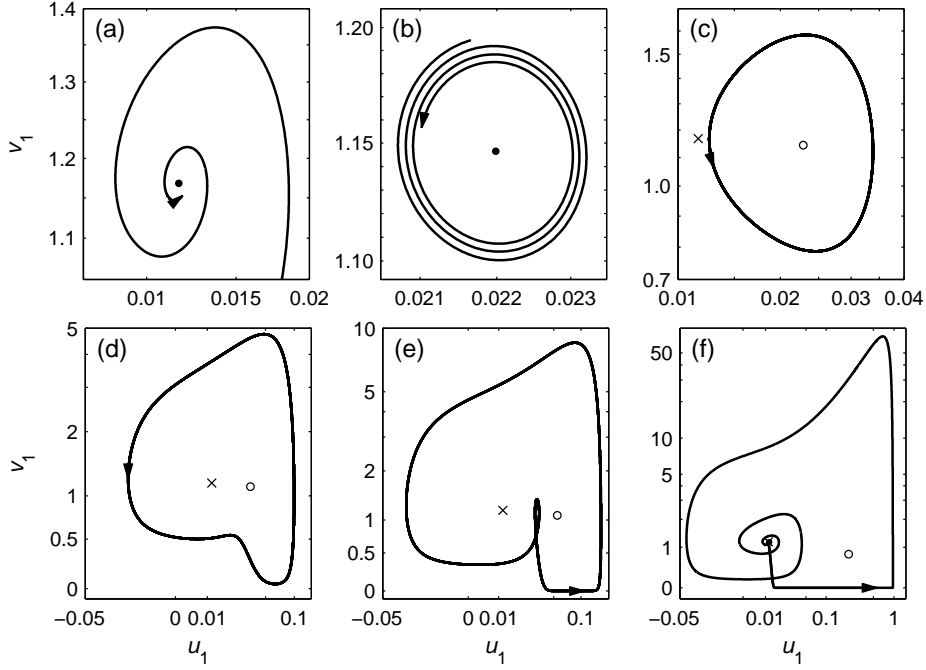}}
\caption{The development of a limit cycle in model \eqref{coupled-uv} in relation to the coupling strength. The projections of phase trajectories and fixed points on $(u_{1},v_{1})$ plane are presented. ``$\bullet$'' and ``$\circ$'' mark the respective stable and unstable internal steady state $F_{12}$; ``$\times$'' stands for the boundary steady state $F_{1}$. (a) $\varkappa_{1,2}=0$: CR pairs are decoupled; each has unique stable steady state. (b) $\varkappa_{1,2}=0.011$: very weak coupling; the spiral winds only algebraically fast. The system is on the verge of the Hopf bifurcation. (c) $\varkappa_{1,2}=0.012$: very weak coupling; the stable limit cycle is just born. (d) $\varkappa_{1,2}=0.025$: weak coupling; nonlinear oscillations. (e) $\varkappa_{1,2}=0.05$: moderate coupling; the limit cycle passes near the basin of the saddle point. (f) $\varkappa_{1,2}=0.6$: moderate coupling; the limit cycle is about to merge with a heteroclinic cycle.}
\label{fig04}
\end{figure}

Fig.~\ref{fig04} plots $(u_{1},v_{1})$ phase planes for coupling strength below and above the Hopf bifurcation. When $\varkappa_{1,2}=0$ (consumers are decoupled) the internal equilibrium is a stable focus (Fig.~\ref{fig04}a). For $\varkappa_{1,2}= 0.011$ (very weak coupling; just below the bifurcation) $F_{12}$ is still a stable focus, though a very gently winding one: the decay is slow (Fig.~\ref{fig04}b). For $\varkappa_{1,2}= 0.012$ (very weak coupling; just above the bifurcation) there is an unstable focus at $F_{12}$ and a stable oval limit cycle of small size representing low-amplitude quasi-harmonic oscillations (Fig.~\ref{fig04}c). On further increasing of coupling strength the limit cycle continuously grows in size and takes an irregular shape indicating nonlinearity of the synchronous oscillations (Fig.~\ref{fig04}d--f).

As a practical matter, the range of very weak coupling not too far away from the Hopf bifurcation, where oscillations are quasi-linear and quasi-harmonic, is of less concern to us than is the range of far more feasible not-too-weak coupling, corresponding to well-developed substantially nonlinear oscillations. We are going to demonstrate that given conditions \eqref{existF12weak}, system \eqref{coupled-uv} exhibits relaxation oscillatory behavior, with the two coupled CR pairs being antiphase locked.

By the assumption, $0<\varepsilon \ll 1$, meaning that system \eqref{coupled-uv} is singularly perturbed. The slow variables are resources, $u_{1}$ and $u_{2}$, and the fast variables are consumers, $v_{1}$ and $v_{2}$. The standard practice of reducing such systems is the adiabatic elimination of the fast variables, when the left-hand side in the fast equation is replaced by zero, thus turning this differential equation into an algebraic equation. It is assumed that the fast variables quickly relax to their momentary equilibrium, quasi-steady-state, values obtained from the algebraic equations, in which the slow variables are treated as parameters. ``Frozen'' slow variables do not move substantially in this short adaptation time of the fast variables. Quasi-steady-state values of the fast variables can then be expressed by values of the slow variables. The fast variables hastily adapt to the motion of the slow variables. The former are entrained by the latter. The utility of quasi-steady-state approximation is that it allows us to reduce the dimension of the system by retaining only slow variables in the model. One has to establish the validity of the adiabatic elimination in each specific case by using the recommendations of the singular perturbation theory \cite{Verhulst:2005}. In particular, Tikhonov's theorem \cite{Tikhonov:1952} requires the quasi-steady state of the fast equations to be stable.

To decompose the full system \eqref{coupled-uv} into fast and slow subsystems, introduce fast time variable $\tau=t/\varepsilon$. Now rescale \eqref{coupled-uv} by replacing $t$ with $\tau\varepsilon$ and, after taking $\varepsilon=0$, it becomes
\begin{subequations}\label{fast}
\begin{align}
u'_{1}& = u'_{2} = 0,\\
v'_{1}& = (u_{1} -\delta v_{1} -\varkappa_{2}v_{2})v_{1},\label{fast-v1}\\
v'_{2}& = (u_{2} -\delta v_{2} -\varkappa_{1}v_{1})v_{2},\label{fast-v2}
\end{align}
\end{subequations}
where prime means differentiation with respect to $\tau$. This is the fast subsystem, where $u_{1}$ and $u_{2}$ are replaced by their initial values and treated as parameters. It yields inner solution, valid for $t = \mathcal{O}(\varepsilon)$.

Setting $\varepsilon=0$ in \eqref{coupled-uv} leads to the slow subsystem
\begin{subequations}\label{slow}
\begin{align}
\dot{u}_{1}& = \gamma_{1} -(u_{1} +1)v_{1} -u_{1},\label{slow-u1}\\
\dot{u}_{2}& = \gamma_{2} -(u_{2} +1)v_{2} -u_{2},\label{slow-u2}\\
0& = (u_{1} -\delta v_{1} -\varkappa_{2}v_{2})v_{1},\label{slow-v1}\\
0& = (u_{2} -\delta v_{2} -\varkappa_{1}v_{1})v_{2},\label{slow-v2}
\end{align}
\end{subequations}
which produces outer solution, valid for $t = \mathcal{O}(1)$. In this singular limit as $\varepsilon \to 0$, the subsystem defines a slow flow on the surface (slow manifold) given by \eqref{slow-v1}--\eqref{slow-v2}. Outer solution is valid for those $u_{1}$ and $u_{2}$, for which the quasi-steady states of the fast subsystem are stable.

We anticipate the dynamics of the full system \eqref{coupled-uv} in its four-dimensional phase space $(u_{1},u_{2},v_{1},v_{2})$ to consist of two typical motions: quickly approaching the slow manifold \eqref{slow-v1}--\eqref{slow-v2}, and slowly sliding over it until a leave point (where the solution disappears) is reached. After that, the representing point may possibly jump to another local solution of \eqref{slow-v1}--\eqref{slow-v2}.

Thus, we ought to find all quasi-steady states of the fast subsystem \eqref{fast}, map the domains of their stability onto the slow phase plane $(u_{1},u_{2})$, and then investigate the dynamics of the slow subsystem \eqref{slow} with piecewise continuous functions.

\begin{figure}[htbp]
\noindent\centering{
\includegraphics[scale=0.50]{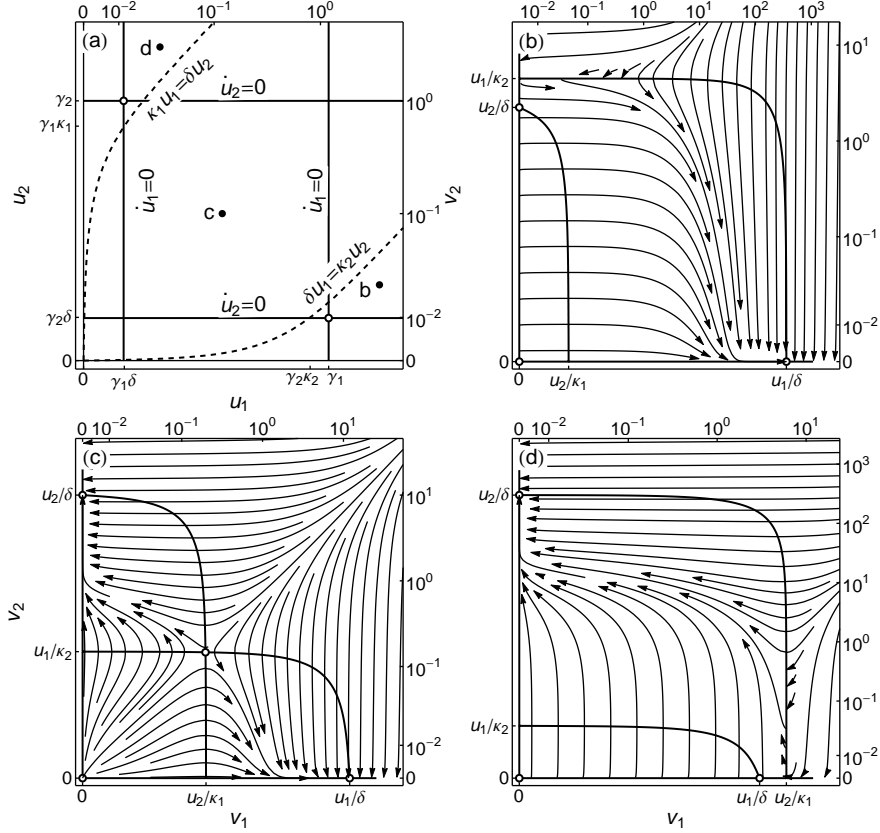}}
\caption{(a) Phase plane of the slow subsystem \eqref{slow} sectored (by dashed lines) into stability domains of the corresponding quasi-steady states of the fast subsystem \eqref{fast}. Both boundary quasi-steady states, each corresponding to the situation when either of the two consumers completely dominates, are stable within the opening of the angle formed by the dashed lines. Lines $u_{1} = \gamma_{1}\delta$ and $u_{2} = \gamma_{2}$ are the respective nullclines $\dot{u}_{1} = 0$ and $\dot{u}_{2} = 0$ of the piecewise subsystem \eqref{slow-piecewise1}. Lines $u_{1} = \gamma_{1}$ and $u_{2} = \gamma_{2}\delta$ mean the same for the piecewise subsystem \eqref{slow-piecewise2}. Intersections of the nullclines (marked by the open circles) are equilibria of the associated piecewise slow subsystems, and they must lie outside the above-mentioned opening to allow for the relaxation oscillations. (b), (c) and (d) are the respective phase portraits of the fast subsystem \eqref{fast} generated by points ``b'', ``c'' and ``d'' in the slow phase plane (a).}
\label{fig05}
\end{figure}

Fast subsystem \eqref{fast}, which is nothing but the conventional LVG model, has four quasi-steady states---three boundary and one interior---denoted by $Q$ (the slow variables are deemed to be frozen):
\begin{subequations}\label{qss}
\begin{alignat}{3}
Q:\qquad &\widetilde{v}_{1}=0,\quad && \widetilde{v}_{2}=0;\label{qssQ}\\
Q_{1}:\qquad &\widetilde{v}_{1}={u_{1}}/{\delta},\quad && \widetilde{v}_{2}=0;\label{qssQ1}\\
Q_{2}:\qquad &\widetilde{v}_{1}=0,\quad && \widetilde{v}_{2}={u_{2}}/{\delta};\label{qssQ2}\\
Q_{12}:\qquad &\widetilde{v}_{1}=\frac{\varkappa_{2}u_{2} -\delta u_{1}}{\varkappa_{1}\varkappa_{2} -\delta^{2}},
\quad && \widetilde{v}_{2}=\frac{\varkappa_{1}u_{1} -\delta u_{2}}{\varkappa_{1}\varkappa_{2} -\delta^{2}}.\label{qssQ12}
\end{alignat}
\end{subequations}
Existence and stability of these quasi-steady states is determined by $(u_{1},u_{2})$---position of a representing point in the phase plane of the slow subsystem, shown in Fig.~\ref{fig05}a.

$Q$, $Q_{1}$ and $Q_{2}$ always exist for all $u_{1}$ and $u_{2}$ from the positive quadrant of the slow phase plane. For not-too-weak coupling, such that $\varkappa_{1}\varkappa_{2}>\delta^{2}$, $Q_{12}$ exists for all $u_{1}$ and $u_{2}$ satisfying the condition $\delta u_{1}/\varkappa_{2} < u_{2} < u_{1}\varkappa_{1}/\delta$, i.\,e. within the opening of the angle formed by lines $\delta u_{1} -\varkappa_{2}u_{2} =0$ and $\varkappa_{1}u_{1} -\delta u_{2} =0$ in Fig.~\ref{fig05}a. The opening shrinks as the coupling strengths get weaker.

Jacobian matrix of the fast subsystem,
\begin{equation}\label{JacobianQ}
\begin{pmatrix}
(u_{1} -2\delta\widetilde{v}_{1} -\varkappa_{2}\widetilde{v}_{2})/\varepsilon & 0 & -\varkappa_{2}\widetilde{v}_{1}/\varepsilon\\
-\varkappa_{1}\widetilde{v}_{2}/\varepsilon & (u_{2} -2\delta\widetilde{v}_{2} -\varkappa_{1}\widetilde{v}_{1})/\varepsilon
\end{pmatrix},
\end{equation}
has the following sets of eigenvalues at \eqref{qss}:
\begin{subequations}\label{eigqss}
\begin{alignat}{5}
\lambda(Q):\qquad&{u_{1}}/{\varepsilon},&&\quad {u_{2}}/{\varepsilon};\label{eigqss-Q}\\
\lambda(Q_{1}):\qquad&-{\varkappa_{1}u_{1}}/\varepsilon\delta +{u_{2}}/{\varepsilon},&&\quad -{u_{1}}/{\varepsilon};\label{eigqss-Q1}\\
\lambda(Q_{2}):\qquad&-{\varkappa_{2}u_{2}}/\varepsilon\delta +{u_{1}}/{\varepsilon},&&\quad -{u_{2}}/{\varepsilon};\label{eigqss-Q2}\\
\lambda(Q_{12}):\qquad&\pm\bigl(\sqrt{u_{1}u_{2}} +\mathcal{O}(\delta)\bigr)/{\varepsilon}.&&\label{eigqss-Q12}
\end{alignat}
\end{subequations}
Based on \eqref{eigqss} one concludes that $Q$ (the origin) is always an unstable node for all $u_{1}$ and $u_{2}$ from the positive quadrant of the slow phase plane. $Q_{1}$ is a stable node for $\delta u_{2} < \varkappa_{1}u_{1}$, i.\,e. below the line $\varkappa_{1}u_{1} -\delta u_{2} = 0$ in Fig.~\ref{fig05}a, otherwise it is a saddle. Similarly, $Q_{2}$ is a stable node for $\delta u_{1} < \varkappa_{2}u_{2}$, i.\,e. above the line $\delta u_{1} -\varkappa_{2}u_{2} = 0$ in the plane of slow variables, otherwise it is a saddle. The interior quasi-steady state, $Q_{12}$, is always a saddle.

The performed typology of fixed points of the fast subsystem \eqref{fast} leads to three qualitatively different phase portraits depicted by Fig.~\ref{fig05}b--d.

Suppose initially $Q_{1}$ is stable and $Q_{2}$ is not. Consumer 1 completely dominates. This corresponds to slow variables $u_{1}$ and $u_{2}$ being somewhere below the line $\delta u_{1} -\varkappa_{2}u_{2} = 0$ of Fig.~\ref{fig05}a. Fast subsystem \eqref{fast} has phase portrait of a type shown in Fig.~\ref{fig05}b. While $u_{1}$ remains much greater than $\gamma_{1}\delta$, the dynamics of the resources (treated as bifurcation parameters in reference to the consumers) is described by a system of two independent equations
\begin{subequations}\label{slow-piecewise1}
\begin{align}
\dot{u}_{1}& = \gamma_{1} -\Bigl(\frac{u_{1} +1}{\delta} +1\Bigr)u_{1},\label{slow-piecewise1-u1}\\
\dot{u}_{2}& = \gamma_{2} -u_{2},\label{slow-piecewise1-u2}
\end{align}
\end{subequations}
which is a piecewise version of the slow subsystem \eqref{slow} for $v_{1}=u_{1}/\delta$ and $v_{2}=0$. System \eqref{slow-piecewise1} has stable steady state
\begin{equation}\label{ss-u1u2-1}
\widehat{u}_{1}^{(1)} =\textstyle\frac{1}{2}(r_{1} -\delta -1) = \gamma_{1}\delta +\mathcal{O}(\delta^{2}),\quad \widehat{u}_{2}^{(1)}=\gamma_{2},
\end{equation}
where $r_{1}=\sqrt{1 +\delta(4\gamma_{1} +2 +\delta)}$. This equilibrium lies in the upper left corner of Fig.~\ref{fig05}a.

While heading to \eqref{ss-u1u2-1}, the trajectory crosses the line $\delta u_{1} -\varkappa_{2}u_{2} = 0$ and enters the domain of bistability of both $Q_{1}$ and $Q_{2}$. Fast subsystem \eqref{fast} takes new phase portrait of a type presented by Fig.~\ref{fig05}c. However the dominance of consumer 1 persists.

Upon introducing the deviations $\xi_{1}$ and $\eta_{2}$ from the respective steady states \eqref{ss-u1u2-1}, the system \eqref{slow-piecewise1} becomes
\begin{subequations}\label{slow-piecewise1-xi-eta}
\begin{align}
\dot{\xi}_{1}& = -\xi_{1}(r_{1} +\xi_{1})/\delta,\label{slow-piecewise1-xi1}\\
\dot{\eta}_{2}& = -\eta_{2},\label{slow-piecewise1-eta2}
\end{align}
\end{subequations}
whence one finds
\begin{equation}\label{sol-xi1-eta2}
\xi_{1}(t)=\frac{r_{1}}{\left(r_{1}/\xi_{1}(0) +1\right)\exp(r_{1}t/\delta) -1}\quad \text{and}\quad \eta_{2}(t)=\eta_{2}(0)\exp(-t).
\end{equation}

It follows from \eqref{sol-xi1-eta2}, that the dynamics of variable $u_{1}$ is faster than that of $u_{2}$ due to small $\delta$. Clearly, the representing point must have relaxed to the vertical line $u_{1}=\gamma_{1}\delta$ well before approaching the horizontal line $u_{2}=\gamma_{2}$. Further developments depend on whether the involved individual CR systems are overdamped or underdamped.

\begin{figure}[htbp]
\noindent\centering{
\includegraphics[scale=0.50]{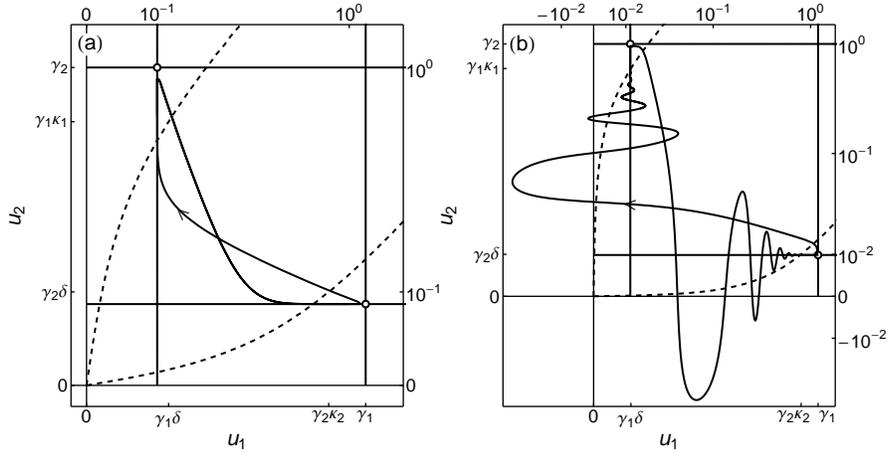}}
\caption{The limit cycle of system \eqref{coupled-uv} projected onto phase plane of the slow variables in the cases of (a) strong second-order damping, and (b) underdamping. The parameters of the model chosen for numerical simulation are $\varepsilon = 0.727273{\times}10^{-3}$, $\gamma_{1} = 1.19375$, $\gamma_{2} = 1$, $\varkappa_{1}=0.5$, $\varkappa_{2}=0.8$, $\delta = 0.1$~(a), and $\delta = 0.01$~(b).}
\label{fig06}
\end{figure}

If $\delta$ is great enough to damp intrinsic oscillations in the constituent CR pairs, the representing point will slide along the nullcline $\dot{u}_{1}=0$ (slow manifold of the system \eqref{slow-piecewise1}), steadily tending to point \eqref{ss-u1u2-1} (Fig.~\ref{fig06}a).

If $\delta$ is small in terms of the condition \eqref{focus-cond-uv}, then the individual CR pairs are underdamped. In the immediate vicinity of $\widehat{u}_{1}^{(1)}$ the division of variables on slow and fast ones loses its meaning and the reduced equation \eqref{slow-piecewise1-u1} is no longer valid. Instead of \eqref{slow-piecewise1-u1} one should write down two equations:
\begin{subequations}\label{constituent-u1v1}
\begin{align}
\dot{u}_{1}& = \gamma_{1} -(u_{1} +1)v_{1} -u_{1},\label{constituent-u1}\\
\varepsilon \dot{v}_{1}& = (u_{1} -\delta v_{1})v_{1}.\label{constituent-v1}
\end{align}
\end{subequations}
However this system is identical to equations \eqref{steady-uv} for an uncoupled CR system and, in essence, describes convergence to a focal point via dying oscillations. In the plane of slow variables $(u_{1},u_{2})$ these oscillations manifest themselves in damped transverse fluctuations superimposed on the independent vertical motion along the nullcline $\dot{u}_{1}=0$ ($u_{1} = \widehat{u}_{1}^{(1)} \approx \gamma_{1}\delta$) toward point \eqref{ss-u1u2-1} (Fig.~\ref{fig06}b).

By virtue of the condition \eqref{existF12weak}, the trajectory has to cross the line $\varkappa_{1}u_{1} -\delta u_{2} =0$ on its way toward the neighborhood of steady state \eqref{ss-u1u2-1}. As soon as this has happened, node $Q_{1}$ in the plane $(v_{1},v_{2})$ will be absorbed by saddle $Q_{12}$. A new phase portrait of the fast subsystem \eqref{fast} takes on the appearance of Fig.~\ref{fig05}d. Consumer 1 rapidly washes out, and the alternative boundary quasi-steady state $Q_{2}$ becomes stable, with consumer 2 dominating.

In terms of the four-dimensional phase space of full system \eqref{coupled-uv}, the representing point is now in the other stable branch of the slow manifold \eqref{slow-v1}--\eqref{slow-v2}. The motion over this alternative branch obeys the piecewise subsystem
\begin{subequations}\label{slow-piecewise2}
\begin{align}
\dot{u}_{1}& =\gamma_{1} -u_{1},\label{slow-piecewise2-u1}\\
\dot{u}_{2}& =\gamma_{2} -\Bigl(\frac{u_{2} +1}{\delta} +1\Bigr)u_{2},\label{slow-piecewise2-u2}
\end{align}
\end{subequations}
with the initial conditions $u_{1}(0) = \widehat{u}_{1}^{(1)} \approx \gamma_{1}\delta$ and $u_{2}(0) = \gamma_{1}\varkappa_{1}$.

The dynamics of \eqref{slow-piecewise2} is basically similar to that of \eqref{slow-piecewise1} analyzed above. System \eqref{slow-piecewise2} has a stable steady state
\begin{equation}\label{ss-u1u2-2}
\widehat{u}_{1}^{(2)} =\gamma_{1},\quad \widehat{u}_{2}^{(2)} =\textstyle\frac{1}{2}(r_{2} -\delta -1) = \gamma_{2}\delta +\mathcal{O}(\delta^{2}),
\end{equation}
where $r_{2}=\sqrt{1 +\delta(4\gamma_{2} +2 +\delta)}$. This equilibrium lies in the lower right corner of Fig.~\ref{fig05}a.

Variable $u_{2}$, being more rapid in comparison to $u_{1}$, quickly enters the neighborhood of the nullcline $\dot{u}_{2} = 0$ given by $u_{2} =\widehat{u}_{2}^{(2)} \approx \gamma_{2}\delta$, and then---depending on the value of $\delta$---finally approaches the nullcline either monotonically (Fig.~\ref{fig06}a) or via damped oscillations according to equations
\begin{subequations}\label{constituent-u2v2}
\begin{align}
\dot{u}_{2}& = \gamma_{2} -(u_{2} +1)v_{2} -u_{2},\label{constituent-u2}\\
\varepsilon \dot{v}_{2}& = (u_{2} -\delta v_{2})v_{2}\label{constituent-v2}
\end{align}
\end{subequations}
(Fig.~\ref{fig06}b). System \eqref{constituent-u2v2} describes underdamped intrinsic oscillations of uncoupled consumer 2 for small $\delta$.

At the same time, $u_{1}$ steadily and independently tends to $\widehat{u}_{1}^{(2)} =\gamma_{1}$. Again, because point \eqref{ss-u1u2-2} is located below the line $\delta u_{1} -\varkappa_{2} u_{2}=0$ (on the strengths of the condition \eqref{existF12weak}), the trajectory would certainly cross that line at a point $(\gamma_{2} \varkappa_{2}, \gamma_{2} \delta)$, whereupon node $Q_{2}$ would be absorbed by saddle $Q_{12}$. The system returns to the first branch of the slow manifold, and thereby the oscillatory cycle gets closed.

\section{Results and discussion}
\begin{figure}[htbp]
\noindent\centering{
\includegraphics[scale=0.55]{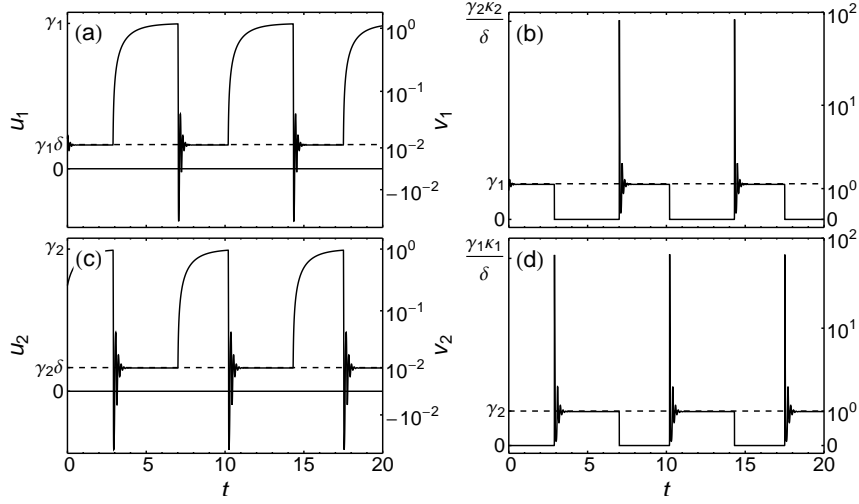}}
\caption{Time profiles of antiphase relaxation oscillations in two coupled CR pairs modeled by \eqref{coupled-uv}. (a) and (b) show the respective resource 1 and consumer 1; (c) and (d) display the respective resource 2 and consumer 2. Numerical values of the parameters are those mentioned in the caption to Fig.~\ref{fig06}b.}
\label{fig07}
\end{figure}

Fig.~\ref{fig07} shows the results of a numerical integration of system \eqref{coupled-uv} for the case of underdamped individual consumer-research pairs. The two coupled communities execute self-sustained relaxation oscillations which are antiphase-locked.

The resources $u_{1}$ and $u_{2}$ demonstrate sawtooth periodic pulses. The oscillation range for the resource levels remains finite and, what is important, it does not depend on the intrinsic second order loss $\delta$ (measuring intraspecific interference).

The times of motion over either branch of the slow manifold \eqref{slow-v1} and \eqref{slow-v2} add up to give a predominant contribution to the period of oscillations, $T$. These times are determined mainly by the dynamics of the slow resource variables $u_{1}$ and $u_{2}$ and, to a zeroth approximation in $\varepsilon$ and $\delta$, can be found as solutions to the equations of motion \eqref{slow-piecewise2-u1} and \eqref{slow-piecewise1-u2} with respective boundary conditions $(0,\gamma_{2} \varkappa_{2})$ and $(0,\gamma_{1} \varkappa_{1})$. In this way one obtains a quite simple estimate for the period:
\begin{equation}\label{period}
T = \int_{0}^{\gamma_{2} \varkappa_{2}} \frac{\mathrm{d}z}{\gamma_{1} -z} +\int_{0}^{\gamma_{1} \varkappa_{1}} \frac{\mathrm{d}z}{\gamma_{2} -z}
= \ln\frac{1}{1 -\varkappa_{2}(\gamma_{2}/\gamma_{1})} +\ln\frac{1}{1 -\varkappa_{1}(\gamma_{1}/\gamma_{2})}.
\end{equation}
It is interesting that, according to \eqref{period}, the period depends on the ratio of the two resource inflows, $\gamma_{1}$ and $\gamma_{2}$, rather than on each of them individually, and does not depend completely on concrete value of $\delta$.

The consumers $v_{1}$ and $v_{2}$ change periodically between extinction and respective constant levels $\gamma_{1}$ and $\gamma_{2}$. Very brief transient from zero to flat nonzero level within each cycle is accompanied by a highly pronounced spiky overshoot. The magnitude of the spike tends to infinity as $\delta \to 0$, in view of \eqref{qssQ1} and \eqref{qssQ2}. Depending on the intensity of intraspecific interference, the overshoot may or may not be followed by a tail of fading high-frequency oscillations, when a consumer variable falls below its steady-state value and then bounce back above, taking some time to settle close to its steady-state value. In signal processing, such a kind of transient oscillations is known as ``ringing''. There is no ringing if the involved CR pair does not oscillate due to significant intraspecific interference. Ringing takes place if intraspecific interference is negligible and therefore the involved CR pair is characterized by underdamped intrinsic oscillations. The ``pitch'' of ringing is just the frequency of these intrinsic oscillations.

One notices that when one consumer is very scanty, the coupled system behaves like an isolated CR pair \eqref{isolated-uv}. Another essential feature of the dynamics is the role of the resource variables in determining when the consumers emerge and wash out. When $u_{1}$, for example, rises above a threshold value (determined by the amount of losses experienced by $v_{1}$) then $v_{1}$ comes into dominance causing $v_{2}$ in turn to disappear. So except for their transient spiking and ringing, the consumer levels, either flat nonzero or essentially zero, are determined by the hysteretic cycling of the respective resources.

Scrutinizing a cycle of consumer oscillations one may distinguish four parts within it:
\begin{compactenum}[1)]
\item $v_{1}$ is essentially zero, while $v_{2}$ is approximately equal to its uncoupled steady-state value, $\gamma_{2}$. $u_{1}$ increases due to its inflow until it overcomes losses for consumer 1;
\item With a sufficient resource stock, $v_{1}$ now emerges. The population exhibits a spike due to the fast time scale of the consumer equations. The sharp increase in population saturates the available resource level, so $u_{1}$ drops. Cross-losses cause $v_{2}$ to wash out;
\item $v_{1}$ and $u_{1}$ relax to quasi-steady-state values, as if there were only one isolated CR pair. $v_{2}$ is essentially zero. $u_{2}$ is increasing, like $u_{1}$ did in part 1;
\item $u_{2}$ surpasses the losses, $v_{2}$ emerges and the subsequent cross-losses cause $v_{1}$ to wash out. The spiking $v_{2}$ also causes a substantial decrease in the available stock of the associated resource. The sequence begins again.
\end{compactenum}

Presenting his famous model Smale remarked that ``it is more difficult to reduce the number of chemicals to two or even three'' \cite{Smale:1974}. As distinct from Smale's example, coupling in our case makes self-sustained synchronous oscillations possible for just two variables.

As we have seen, phase trajectory of the system constantly moves from the neighborhood of unstable boundary equilibrium $F_{1}$ where only consumer 1 is present, to the neighborhood of $F_{2}$ where consumer 2 completely dominates, back to $F_{1}$, and so on in cyclic alternation. This kind of trajectory was termed ``heteroclinic cycle'' by Kirlinger \cite{Kirlinger:1986}. A heteroclinic cycle occurs when the outflow (unstable manifold) from one saddle point is directly connected to the inflow (stable manifold) of another saddle point, and vice versa. It is closely related to another notion of the nonlinear dynamics, a homoclinic cycle, which emerges when the unstable and the stable manifolds of the same saddle coincide and form a closed loop.

\begin{figure}[htbp]
\noindent\centering{
\includegraphics[scale=0.6]{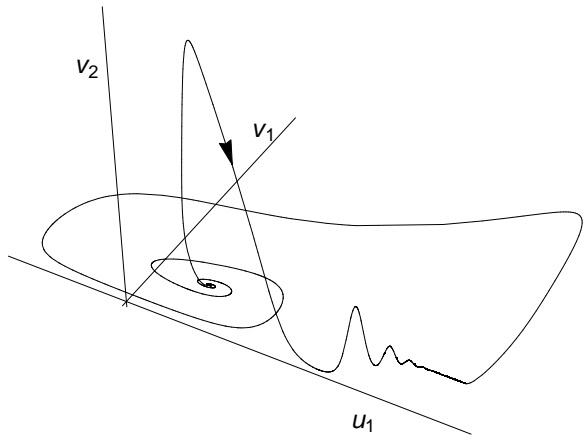}}
\caption{A 3D-projection of the limit cycle in system \eqref{coupled-uv} for parameters chosen in a neighborhood of the heteroclinic cycle. See caption to Fig.~\ref{fig06}b for the parameters.}
\label{fig08}
\end{figure}

Homo- and heteroclinic cycles are not robust structures in the sense that infinitesimally small change of system parameters destroy them. However in the practical sense, any limit cycle passing in close proximity to saddle points will be indistinguishable from a heteroclinic cycle (Fig.~\ref{fig08}). The only difference is strict periodicity, although the period of the limit cycle in a neighborhood of the heteroclinic cycle may be long. Besides, at the threshold of homo-/heteroclinic bifurcation the period is susceptible to external noise.

In the context of our model, as coupling becomes stronger, the stable limit cycle swells and passes closer and closer to boundary fixed points which are node-saddles or focus-saddles (Fig.~\ref{fig04}d--f). Depending on the interplay between the parameters, eventually it may bang into one or both of these equilibria creating either a homoclinic or heteroclinic cycle, respectively. This corresponds to $\gamma_{2}/\gamma_{1}=\varkappa_{1}$ and $\gamma_{1}/\gamma_{2}=\varkappa_{2}$. On further increasing the coupling, the saddle connection breaks and the loop is destroyed.

It is worth noting that heteroclinic cycles were first found by May and Leonard \cite{May:1975} in a classical LVG system with competing three species. However in their model the cycle is not truly periodic: as time goes on, the system tends to stay in the neighborhood of any one boundary equilibrium ever longer, so that the ``total time spent in completing one cycle is likewise proportional to the length of time the system has been running.'' Moreover, May and Leonard state that ``the phenomenon clearly requires at least three competitors, which is why it cannot occur in models with two competitors.'' This statement is echoed by Vandermeer \cite{Vandermeer:2011} who extended their theory on higher dimensions: ``It appears to be the case that all cases of an odd number of species follow this basic pattern, whereas all cases of even number of species result in extinction of half of the components, leaving the other half living independently at their carrying capacities.'' In view of our results, the above conclusion is by far and away true providing one stays within the framework of classical LVG equations, which in fact imply a high rapidity of the resource dynamics. In our model of just two competitors the slowness of the resource relative to the consumer is essential for the oscillations to occur, because it provides the necessary inertia to the system.

Physically, our model is most likely feasible because it is based on the well-established rate equations \eqref{rateeqns-p} for semiconductor lasers, and therefore should be considered as a model of anti-phase synchronization of two lasers via their loss-coupling.

Ecologically, the feasibility of the model is tightly bound to justification of the adopted time hierarchy in system \eqref{coupled-uv}. Time scales are usually inverted in ecosystems, as opposed to lasers, the most common case being rapid consumption of food by species. However it seems reasonable to propose that our model may describe the first level of an ecosystem, at which the consumers are autotrophs and the resources are mineral nutrients. The ability to exploit different substrates leads to a possibility of stable coexistence of different organisms descending from the common ancestor. Divergent evolution is just the formation of new species: due to mutations two populations emerge with the same genetic code but having proteins able to process different substrates. Providing the environmental conditions are quite stable on the evolutionary timescale, the inflows of inorganic substrates from the surroundings may be considered constant and the washout time of a substrate may occur much longer than the life expectancy of a species (recall the definition of $\varepsilon$ from \eqref{scaling}).

\section{Conclusion} We proposed a model of two CR pairs linked by interspecific interference competition. When uncoupled, an individual CR pair has a unique stable steady state and does not admit periodic solutions. If intraspecific interference within the species is strong enough, the equilibrium is nonoscillatory (stable node), otherwise the steadying occurs by decaying oscillations (stable focus).

When coupled, the model behaves differently at strong and weak competitive interaction between the consumers. When coupling is strong, one of the consumers wins. Which consumer wins or loses depends critically on the relative intensities of the resource inflows and coupling strengths. In the case of bistability, when the system acts like a bistable multivibrator (flip-flop circuit), the winner may be determined by the initial conditions. Any static coexistence of competing consumers is not possible.

When coupling is moderately weak, the model reveals low-frequency antiphase relaxation oscillations represented by a continuous flow of rectangular pulses. The system works as an astable multivibrator continually switching between its two states, neither of which is stable. The consumers cannot coexist even dynamically: in each of two alternating states one consumer completely dominates and the other is on the verge of extinction. The most intriguing feature of the model is that each of the participating CR pairs taken separately does not oscillate; both communities are completely quiescent, however, in interaction, when coupled in a nonlinear way, the resulting system turns into a relaxation oscillator.

One way or the other, it is believed that the proposed model for coupling-induced oscillations in nonoscillatory CR pairs can be considered as a minimal in that class of population-dynamical systems and its mechanism can be applied to networks with large numbers of nonoscillatory elements and complex architecture.

\section*{Statement on the absence of conflict of interests}
The author declares that there is no conflict of interests regarding the publication of this paper.

\setcounter{section}{0}
\setcounter{equation}{0}
\renewcommand{\theequation}{\thesection.\arabic{equation}}
\setcounter{table}{0}
\renewcommand{\thetable}{\thesection.\arabic{table}}

\appendix
\section*{Appendix. Derivation of the coupled CR equations}
\renewcommand{\thesection}{A}

\subsection{Ecological perspective}\label{app1}
Of all types of interactions between individuals of the same population (intraspecific interactions) or individuals of different populations (interspecific interactions) of the same trophic level competition is most commonly encountered. In a broad sense, competition takes place when each species (individual) has an inhibiting effect on the growth of the other species (individual). An inhibiting effect should be understood to mean either an increase in the death rate or a decrease in the birth rate.

Consider the famous CR equations proposed by MacArthur \cite{MacArthur:1970, Chesson:1990}:
\begin{subequations}\label{CRMac}
\begin{alignat}{3}
\dot{x}_{j}& = \Bigl(r_{j}(1 -{x_{j}}/{K_{j}}) -\sum_{i=1}^{n}c_{ij}y_{i}\Bigr)x_{j},\qquad & j&=1,\dotsc,m,\label{CRMac-x}\\
\dot{y}_{i}& = \Bigl(\sum_{j=1}^{m}c_{ij}w_{j}x_{j} -b_{i}\Bigr)y_{i},\qquad & i&=1,\dotsc,n.\label{CRMac-y}
\end{alignat}
\end{subequations}
Here dots indicate differentiation with respect to time $t$, $x_{j}$ represents the total biomass of $j$th resource (prey), $y_{i}$ stands for the total biomass of $i$th consumer (predator) species, the constant $r_{j}$ defines the growth rate of $j$th resource, $K_{j}$ is the carrying capacity of $j$th resource, $c_{ij}$ is the rate of uptake of a unit of $j$th resource by each individual of $i$th consumer population, $w_{j}^{-1}$ is the conversion efficiency parameter representing an amount of $j$th resource an individual of $i$th consumer population must consume in order to produce a single new individual of that species, $b_{i}$ is the loss rate of $i$th consumer due to either natural death or emigration. All parameters in \eqref{CRMac} are nonnegative.

MacArthur assumed population dynamics of the resources to be much faster than that of the consumers which enabled him to approximate $x_{j}$ in \eqref{CRMac-y} by its quasi-steady-state value derived by setting the right-hand side of \eqref{CRMac-x} to zero. As a result, he succeeded to reduce slow-scale equation \eqref{CRMac-y} to the well-known LVG model \cite{Gause:1935}
\begin{subequations}\label{VG}
\begin{alignat}{5}
\dot{y}_{i}&= \Bigl(k_{i} -\sum_{s=1}^{n}a_{is}y_{s}\Bigr)y_{i},\qquad & i&=1,\dotsc,n,\label{VG-y}\\
\intertext{where}
a_{is}&= \sum_{j=1}^{m}c_{ij}c_{sj}(w_{j}K_{j}/r_{j}),\qquad & i&=1,\dotsc,n;\quad s=1,\dotsc,n,\label{VG-a}\\
\intertext{and}
k_{i}&= \sum_{j=1}^{m}c_{ij}w_{j}K_{j} -b_{i},\qquad & i&=1,\dotsc,n.\label{VG-k}
\end{alignat}
\end{subequations}

More recently, such an asymptotic reduction has also been carried out for a model of competition where species (with continuous trait) consume the common resource that is constantly supplied, under the assumption of a very fast dynamics for the supply of the resource and a fast dynamics for death and uptake rates \cite{Mirrahimi:2013}.

CR model \eqref{CRMac} assumes that competition within and between consumer species is purely exploitative: individuals and populations interact through utilizing (or occupying) common resource that is in short supply. Quite on the contrary, LVG model \eqref{VG} describes competition strictly phenomenologically, as direct interference where consumers experience harm attributed to their mutual presence in a habitat (e.g. through aggressive behavior). However we have to stress that neither MacArthur's reduction claims that interference competition entirely results from ``more fundamental'' trophic competition, nor it urges us to hastily consider direct competition as some derived concept. What it actually states is that at slow-time scale associated with dynamics of the consumers, the effects of exploitation competition are indistinguishable from those of interference competition. And at slow-time scale, coefficients $a_{is}$ of \eqref{VG-a} merely add to interference coefficients $a'_{is}$ which are to be present primordially in \eqref{CRMac-y}.

Most mathematical models dealing with coupled CR pairs or multilevel trophic chains ignore contributions of intraspecific and interspecific interference. Indeed, the empirical data like \cite{Devetter:2008} do indicate that $a'_{ij}$ may be negligible in comparison with $a_{is}$. Still, works advocating the explicit account for direct interference show that incorporation of self-limitation and cross-limitation terms in the equations at the consumers' level can provide for the stable coexistence of many species on few resources \cite[p. 31]{Bazykin:1998}, \cite{Kuang:2003}.

Moreover, if we are to assume dynamics of the resources to be much slower than that of the consumers, it is likely that we have to retain interference competition terms in all equations \eqref{CRMac-y}.

Consider the following modification of \eqref{CRMac} representing coupled two-consumer, two-resource equations:
\begin{subequations}\label{CRcoupled}
\begin{align}
\dot{x}_{1}& = p_{1} -(c_{1} y_{1} +q_{1}) x_{1},\label{CRcoupled-x1}\\
\dot{x}_{2}& = p_{2} -(c_{2} y_{2} +q_{2}) x_{2},\label{CRcoupled-x2}\\
\dot{y}_{1}& = (c_{1} w_{1} x_{1} -b_{1} -d_{1} y_{1} -h_{2}y_{2})y_{1},\label{CRcoupled-y1}\\
\dot{y}_{2}& = (c_{2} w_{2} x_{2} -b_{2} -d_{2} y_{2} -h_{1}y_{1})y_{2}.\label{CRcoupled-y2}
\end{align}
\end{subequations}

Instead of the logistic mode of resource supply, as is the case in MacArthur's model, our model is based on so-called ``equable'' mode of resource exploitation \cite{StewartLevin:1973}, by which the quantities of available resources are held constant by a continuous-flow system. According to  \eqref{CRcoupled-x1} and \eqref{CRcoupled-x2}, a constant concentration of $j$th resource ($j=1,2$) flows into a defined volume with the rate $p_{j}$ while unused resource flows out with the per capita rate $q_{j}$, in much the same manner as in a chemostat \cite{Herbert:1956}.

In more exact terms, the true chemostat model for one substrate and one species looks as follows:
\begin{equation}\label{chemostat}
\begin{split}
\dot{x}& = D(x_{0} -x) -\frac{\mu x y}{K_{x} +x},\\
\dot{y}& = \Bigl(\frac{w \mu x}{K_{x} +x} -D\Bigr)y,
\end{split}
\end{equation}
where the rate of substrate uptake is expressed by the Monod formula $\mu x y/(K_{x} +x)$, $K_{x}$ is a saturation constant numerically equal to the substrate concentration at which the uptake rate is half the maximum, $D$ is the dilution rate defined as the rate of flow of medium over the volume of the bioreactor, and $x_{0}$ is an input concentration of the substrate.

Model \eqref{chemostat} turns into an uncoupled version of \eqref{CRcoupled} if we put $p=Dx_{0}$, $q=b=D$, and assume $K_{x} \gg x$, so that $\mu x/(K_{x} +x) \approx c x$, where $c=\mu/K_{x}$.

In natural conditions, the equable modes of feeding, for instance, can be found on the first trophic level of ecosystem, among autotrophs.

Besides, in \eqref{CRcoupled-y1} and \eqref{CRcoupled-y2} intraspecific competition strength $d_{i}$ ($i=1,2$) measures direct interference of individuals within $i$th consumer population with each other resulting in an additional per capita loss rate $d_{i} y_{i}$; interspecific competition strength $h_{s}$ ($s=1,2$; $s\neq i$) quantifies direct interference effect from $s$th consumer on $i$th consumer resulting in an additional per capita loss rate, $h_{s}y_{s}$, of the latter.

Equations \eqref{CRcoupled} contain two important assumptions. First, they assume that the resources are noninteractive. On higher trophic levels, however, resources may interact and the possibility of competition among the resources was originally pointed out by Lynch \cite{Lynch:1978}. Since then, a whole series of theoretical papers have been published on two-predator, two-prey systems with interference competition between two self-reproducting prey species based on the Lotka--Volterra equations. Specifically, Kirlinger \cite{Kirlinger:1986} describes the model in which each predator specializes on one prey only, while Xiang and Song \cite{Xiang:2006} treat a similar model in which each predator is allowed to feed on both prey.

As seen from \eqref{CRcoupled-x1} and \eqref{CRcoupled-x2}, there is no intraspecific interference competition within the resource populations either. Yet the resource level would remain finite even in the absence of the consumer.

A second assumption of our equations is that the consumers interact only directly, through interference competition, and cannot compete through their use of resources, as each consumer specializes on one resource only. The theory of pure trophic competition in equable models has been developed in the works \cite{StewartLevin:1973, Tilman:1982}.

Intraspecific interference competition is allowed within the consumers as well. Even though the available amount of any resource happened to be of a constant level, the population size of the associated consumer would remain finite due to self-limitation caused by direct intraspecific interference.

The novelty of model \eqref{CRcoupled} is that it considers time hierarchy of MacArthur's CR equations to be reversed by assuming dynamics of the consumers to be much faster than that of the involved resources and articulates the importance of direct competition mechanisms within the framework of this assumption.

Upon the scaling
\begin{equation}\label{scaling}
\begin{gathered}
u_{1} = \frac{c_{1}w_{1}x_{1}}{b_{1}} -1,\quad u_{2} = \frac{c_{2}w_{2}x_{2}}{b_{2}} -1,\quad
v_{1} = \frac{c_{1}y_{1}}{q_{1}},\quad v_{2} = \frac{c_{2}y_{2}}{q_{2}},\\
\gamma_{1} =\frac{c_{1}p_{1}w_{1}}{b_{1}q_{1}} -1,\quad \gamma_{2} = \frac{c_{2}p_{2}w_{2}}{b_{2}q_{2}} -1,\quad
\delta_{1} = \frac{d_{1}q_{1}}{b_{1}c_{1}},\quad \delta_{2} = \frac{d_{2}q_{2}}{b_{2}c_{2}},\\
\varkappa_{1} = \frac{h_{1}q_{1}}{b_{2}c_{1}},\quad \varkappa_{2} = \frac{h_{2}q_{2}}{b_{1}c_{2}},\quad \beta = \frac{q_{1}}{q_{2}},\quad
\varepsilon_{1} = \frac{q_{1}}{b_{1}},\quad \varepsilon_{2} = \frac{q_{2}}{b_{2}},\quad t' = q_{1}t
\end{gathered}
\end{equation}
equations \eqref{CRcoupled} take the following nondimensional form:
\begin{subequations}\label{CRnondim}
\begin{align}
\dot{u}_{1}& =\gamma_{1} -u_{1}v_{1} -u_{1} -v_{1},\label{CRnondim-u1}\\
\beta\dot{u}_{2}& =\gamma_{2} -u_{2}v_{2} -u_{2} -v_{2},\label{CRnondim-u2}\\
\varepsilon_{1}\dot{v}_{1}& =(u_{1} -\delta_{1}v_{1} -\varkappa_{2}v_{2})v_{1},\label{CRnondim-v1}\\
\varepsilon_{2}\dot{v}_{2}& =(u_{2} -\delta_{2}v_{2} -\varkappa_{1}v_{1})v_{2}.\label{CRnondim-v2}
\end{align}
\end{subequations}
Note that in \eqref{CRnondim} dots mean differentiation with respect to nondimensional ``slow'' timescale variable $t'$, as defined by \eqref{scaling}.

The parameters $\beta^{-1}$, $\varepsilon_{1}^{-1}$ and $\varepsilon_{2}^{-1}$ reflect the rapidity of the dynamics of $u_{2}$, $v_{1}$ and $v_{2}$ with reference to that of $u_{1}$. It is assumed that $\beta=\mathcal{O}(1)$ and $\varepsilon_{1,2}\ll 1$.

In studying the effect of coupling, the parameters of interest are obviously the coupling strengths, $\varkappa_{1}$ and $\varkappa_{2}$. The parameters of interest are also those which characterize the difference between the states of the uncoupled systems. The resource income rates $\gamma_{1}$ and $\gamma_{2}$ are used as the control parameters that distinguish the relative base states of the two systems.

For the sake of simplicity but without any loss of generality, we set $\beta=1$, $\varepsilon_{1}=\varepsilon_{2}=\varepsilon$ and $\delta_{1}=\delta_{2}=\delta$, and also drop the prime at $t$, to obtain \eqref{coupled-uv}.

\subsection{Laser dynamics perspective}\label{app2}
Laser rate equations originally proposed by Statz and deMars \cite{Statz-deMars:1960} are differential equations that relate two quantities: injected carrier density ($n$) and photon density ($p$). For a single-mode semiconductor laser, these equations take the form \cite[ch. 6]{Agrawal:1993}
\begin{subequations}\label{rateeqns}
\begin{align}
\dot{n}&= \frac{J}{q d} -\Gamma v_{g}a(n -n_{0})p -\frac{n}{\tau_{e}},\label{rateeqns-n}\\
\dot{p}&= \Gamma v_{g}a(n -n_{0})p -\frac{p}{\tau_{p}},\label{rateeqns-p}
\end{align}
\end{subequations}
where dots mean differentiation with respect to time $t$, $J$ is the injection current density (pump parameter), $q$ is the magnitude of the electron charge, $d$ is the active-layer thickness, $\Gamma$ is the confinement factor accounting for the fraction of the light power contained in the active region, $v_{g}$ is the group velocity of light that can be expressed through the speed of light in vacuum ($c$) and the group refractive index of the dispersive semiconductor material ($\mu_{g}$) as $v_{g}=c/\mu_{g}$, $a$ is the gain coefficient, $n_{0}$ is the carrier density at transparency corresponding to the onset of population inversion, $\tau_{e}$ is the lifetime of the electrons in the conduction band before being lost by escape from the active region, $\tau_{p}$ is the lifetime of photons inside the cavity before going out of the cavity or being absorbed inside the cavity. In \eqref{rateeqns-p} the contribution of spontaneous emission is neglected.

Typical parameter values for a semiconductor laser (mostly borrowed from \cite[p. 238]{Agrawal:1993}) are given in Table \ref{tabA1}. These numerical values are used in the calculations throughout the present paper, unless otherwise noted.
\begin{table}[htbp]
\caption{Laser parameters used for numerical simulations. The values in parentheses stand for associated nondimensional quantities. }
\begin{center}\footnotesize
\begin{tabular}{lll}
\toprule
\multicolumn{2}{c}{Quantity} \\
\cmidrule(r){1-2}
Notation           & Meaning                      & Value \\
\midrule
$J$\ ($\gamma$)    & pump current density         & $5\times 10^{3}\ \mathrm{A}/\mathrm{cm}^{2}$\ (1.19375) \\%
$d$                & active layer thickness       & $2\times 10^{-5}\ \mathrm{cm}$ \\%
$\Gamma$           & confinement factor           & $0.3$ \\%
$\mu_{g}$          & group refraction index       & $4$ \\%
$a$                & differential gain coefficient& $2.5\times 10^{-16}\ \mathrm{cm}^{2}$ \\%
$n_{0}$            & carrier density at transparency& $1\times 10^{18}\ \mathrm{cm}^{-3}$ \\%
$\tau_{e}$         & carrier lifetime             & $2.2\times 10^{-9}\ \mathrm{s}$ \\%
$\tau_{p}$         & photon loss time             & $1.6\times 10^{-12}\ \mathrm{s}$ \\%
$\varepsilon$      & $\tau_{p}/\tau_{e}$          & $0.727273\times 10^{-3}$ \\%
$D$\ ($\delta$)    & intrinsic second order loss  & $0.773\times 10^{-5}\ \mathrm{cm}^{3}/\mathrm{s}$\ (0.01) \\%
\bottomrule
\end{tabular}
\end{center}
\label{tabA1}
\end{table}

Consider two (not necessarily identical) lasers of type \eqref{rateeqns} and introduce additional intensity-dependent losses such that each laser, $i$, of the two has a total loss represented by the sum of the constant loss, $1/{\tau_{p}}_{i}$, plus the loss proportional to its own intensity, $D_{i}p_{i}$, plus the loss proportional to the intensity of the other laser, $h_{j}p_{j}$:
\begin{subequations}\label{coupledlas}
\begin{align}
\dot{n}_{1}&= \frac{J_{1}}{q d_{1}} -\Gamma_{1} v_{g1}a_{1}(n_{1} -n_{01})p_{1} -\frac{n_{1}}{{\tau_{e}}_{1}},\label{coupledlas-n1}\\
\dot{n}_{2}&= \frac{J_{2}}{q d_{2}} -\Gamma_{2} v_{g2}a_{2}(n_{2} -n_{02})p_{2} -\frac{n_{2}}{{\tau_{e}}_{2}},\label{coupledlas-n2}\\
\dot{p}_{1}&= \Bigl(\Gamma_{1} v_{g1}a_{1}(n_{1} -n_{01}) -\frac{1}{{\tau_{p}}_{1}} -D_{1}p_{1} -h_{2}p_{2}\Bigr)p_{1},\label{coupledlas-p1}\\
\dot{p}_{2}&= \Bigl(\Gamma_{2} v_{g2}a_{2}(n_{2} -n_{02}) -\frac{1}{{\tau_{p}}_{2}} -D_{2}p_{2} -h_{1}p_{1}\Bigr)p_{2},\label{coupledlas-p2}
\end{align}
\end{subequations}
where $D_{i}$ ($i=1,2$) is the second order loss constant of the isolated $i$th laser and $h_{i}$ is the coupling strength measuring the cross-loss effect of $i$th laser on $j$th laser.

Thus according to \eqref{coupledlas}, two lasers happen to be cross-coupled through their resonators, so that each of them can modulate the cavity loss of the other. Technically, intensity-dependent intrinsic and cross-losses may be implemented, for example, using an intracavity electro-optic modulator fed by a current proportional to the output power \cite{Ghosh:2010, Tian:2012}.

We will consider the pump currents $J_{1}$ and $J_{2}$, and the coupling strengths $h_{1}$ and $h_{2}$, as free parameters of the model. For the present, we cannot judge with any confidence the numerical value of $D_{1,2}$, however, as it is demonstrated in the main body of the paper, the exact value of the intrinsic second order loss is not all that critical and does not affect principal results of our analysis, providing that this parameter is small in a sense. For the purposes of model calculations, we adopt $D_{1,2}$ to be somewhat less than $4\times 10^{-5}\ \mathrm{cm}^{3}/\mathrm{s}$.

Presumably, Hofelich-Abate and Hofelich \cite{Hofelich-Abate:1969} were first to introduce (in general form) an intensity-dependent self-limitation term in the photon-density rate equation. Later, that has been done in an explicit form of the second order loss \cite{Kennedy:1974, Kellou:1998}. Those and subsequent studies \cite{Li:2012, El-Amili:2013} showed efficient damping of relaxation oscillations in the presence of intensity-dependent losses.

As to the formal analysis of laser coupling via intensity-loss cross-modulation, very few attempts have been done so far to consider such a mechanism---e.\,g. \cite{Nguyen:1999} and \cite{Mustafin:2006}. The former work, however, does not treat intrinsic second order losses.

The works which address loss-coupled modes of a single laser rather than loss-coupled single-mode lasers are much more plentiful and varied, and we refer the reader for the reviews in \cite[ch. 8]{Mandel:1997} and \cite[ch. 12]{Erneux:2010}. In his seminal paper \cite{Baer:1986}, Baer studied multimode regime of Nd:YAG (neodymium-doped yttrium aluminum garnet) laser with the intracavity-doubling KTP (potassium titynal phosphate) crystal both experimentally and numerically, and proposed the following rate equations for two coupled longitudinal modes:
\begin{subequations}\label{baer}
\begin{align}
\tau_{f}\dot{G}_{1}&= G_{1}^{0} -(\beta I_{1} +\beta_{12} I_{2} +1)G_{1},\label{baer-G1}\\
\tau_{f}\dot{G}_{2}&= G_{2}^{0} -(\beta I_{2} +\beta_{21} I_{1} +1)G_{2},\label{baer-G2}\\
\tau_{e}\dot{I}_{1}&= (G_{1} -\alpha_{1} -\varepsilon I_{1} -2\varepsilon I_{2})I_{1},\label{baer-I1}\\
\tau_{e}\dot{I}_{2}&= (G_{2} -\alpha_{2} -\varepsilon I_{2} -2\varepsilon I_{1})I_{2},\label{baer-I2}
\end{align}
\end{subequations}
where $G_{i}$ and $I_{i}$ ($i=1,2$) are the respective gain and intensity of $i$th mode, $\tau_{f}$ is the fluorescence lifetime, $G_{i}^{0}$ is the small-signal gain (pump parameter) for $i$th mode, $\beta$ is the saturation parameter which determines how strongly the intensity depletes the available gain, $\beta_{12}=\beta_{21}$ is the cross-saturation parameter for modes 1 and 2, $\tau_{e}$ is the cavity round-trip time, $\alpha_{i}$ is the loss of $i$th mode, $\varepsilon$ is the nonlinear coupling coefficient, which models the intracavity-doubling crystal as an intensity-dependent loss in the laser resonator.

The last two terms in \eqref{baer-I1} and \eqref{baer-I2} represent second order losses that are due to intracavity second-harmonic generation and sum-frequency generation, respectively. Numerical calculations revealed antiphase synchronous oscillations in \eqref{baer}. In antiphase state, either mode has precisely the same time profile being shifted by $1/2$ of a period from its counterpart. This type of dynamics was later observed in a multimode $\mathrm{Nd}^{3+}$:YAG laser with intracavity doubling crystal \cite{Wiesenfeld:1990}. For $n$ antiphase oscillators, the phase shift between any nearest neighbors is $1/n$ of a period.

At $\beta_{12}=\beta_{21}=0$, system \eqref{baer} is essentially identical to \eqref{coupledlas}, except that intensity-dependent intrinsic and cross-losses in \eqref{coupledlas-p1} and \eqref{coupledlas-p2} are allowed to be independent.

The scaling
\begin{equation}\label{scaling-baer}
\begin{gathered}
u_{1} = \Gamma_{1}v_{g1}a_{1}{\tau_{p}}_{1}(n_{1} -n_{01}) -1,\quad
u_{2} = \Gamma_{2}v_{g2}a_{2}{\tau_{p}}_{2}(n_{2} -n_{02}) -1,\\
v_{1} = \Gamma_{1}v_{g1}a_{1}{\tau_{e}}_{1}p_{1},\quad
v_{2} = \Gamma_{2}v_{g2}a_{2}{\tau_{e}}_{2}p_{2},\\
\gamma_{1} = \Gamma_{1}v_{g1}a_{1}{\tau_{p}}_{1}\Bigl(\frac{J_{1}{\tau_{e}}_{1}}{qd_{1}} -n_{01}\Bigr) -1,\quad
\gamma_{2} = \Gamma_{2}v_{g2}a_{2}{\tau_{p}}_{2}\Bigl(\frac{J_{2}{\tau_{e}}_{2}}{qd_{2}} -n_{02}\Bigr) -1,\\
\delta_{1} = \frac{D_{1}{\tau_{p}}_{1}}{\Gamma_{1}v_{g1}a_{1}{\tau_{e}}_{1}},\quad
\delta_{2} = \frac{D_{2}{\tau_{p}}_{2}}{\Gamma_{2}v_{g2}a_{2}{\tau_{e}}_{2}},\quad
\varkappa_{1} = \frac{h_{1}{\tau_{p}}_{2}}{\Gamma_{1}v_{g1}a_{1}{\tau_{e}}_{1}},\\
\varkappa_{2} = \frac{h_{2}{\tau_{p}}_{1}}{\Gamma_{2}v_{g2}a_{2}{\tau_{e}}_{2}},\quad
\beta = \frac{{\tau_{e}}_{2}}{{\tau_{e}}_{1}},\quad
\varepsilon_{1} = \frac{{\tau_{p}}_{1}}{{\tau_{e}}_{1}},\quad
\varepsilon_{2} = \frac{{\tau_{p}}_{2}}{{\tau_{e}}_{2}},\quad
t' =\frac{t}{{\tau_{e}}_{1}}
\end{gathered}
\end{equation}
turns \eqref{coupledlas} into nondimensional form \eqref{CRnondim}. Since the lasers are made of the same material, we can put $d_{1}=d_{2}$, $\Gamma_{1}=\Gamma_{2}$, $v_{g1}=v_{g2}$, $a_{1}=a_{2}$, $n_{01}=n_{02}$, ${\tau_{e}}_{1}={\tau_{e}}_{2}$, ${\tau_{p}}_{1}={\tau_{p}}_{2}$, and $D_{1}=D_{2}$, whence $\beta=1$, $\varepsilon_{1}=\varepsilon_{2}=\varepsilon$ and $\delta_{1}=\delta_{2}=\delta$, and we arrive, dropping the prime at $t$, at the set of equations \eqref{coupled-uv}.


\end{document}